\documentclass[twocolumn,showpacs,preprintnumbers,amsmath,amssymb,prb,superscriptaddress]{revtex4-1}

\usepackage{multirow}
\usepackage{amsfonts}
\usepackage{amsopn}
\usepackage[english]{babel}
\usepackage[T1]{fontenc}
\usepackage{times}
\usepackage{mathrsfs}
\usepackage{graphicx}
\usepackage{dcolumn}
\usepackage{bm}
\usepackage[colorlinks,bookmarks=true,citecolor=blue,linkcolor=red,urlcolor=blue]{hyperref}

\begin{document}
\author{Zhao Liu}
\affiliation{Dahlem Center for Complex Quantum Systems and Institut f\"ur Theoretische Physik, Freie Universit\"at Berlin, Arnimallee 14, 14195 Berlin, Germany}
\author{R. N. Bhatt}
\affiliation{Department of Electrical Engineering, Princeton University, Princeton, New Jersey 08544, USA}

\title{Evolution of quantum entanglement with disorder in fractional quantum Hall liquids}
\date{\today}

\begin{abstract}
We present a detailed study of the ground-state entanglement in disordered fractional quantum Hall liquids. We consider electrons at various filling fractions $f$ in the lowest Landau level, with Coulomb interactions. At $f=1/3,1/5$, and $2/5$ where an incompressible ground-state manifold exists at zero disorder, we observe a pronounced minimum in the derivative of entanglement entropy with respect to disorder. At each filling, the position of this minimum is stable against increasing system size, but its magnitude grows monotonically and appears to diverge in the thermodynamic limit. We consider this behavior of the entropy derivative as a compelling signal of the expected disorder-driven phase transition from a topological fractional quantum Hall phase to an insulating phase. On the contrary, at $f=1/2$ where a compressible composite fermion sea is present at zero disorder, the entropy derivative exhibits much greater, almost chaotic, finite-size effects, without a clear phase transition signal for system sizes within our exact diagonalization limit. However, the dependence of entanglement entropy with system size changes with increasing disorder, consistent with the expectation of a phase transition from a composite fermion sea to an insulator. Finally, we consider $f=1/7$ where compressible Wigner crystals are quite competitive at zero disorder, and analyze the level statistics of entanglement spectrum at $f=1/3$.
\end{abstract}

\maketitle

\section{Introduction}
The rich physics of two-dimensional (2D) electron systems subject to a strong perpendicular magnetic field in the fractional quantum Hall (FQH) regime has attracted intensive interest. In particular, studying the FQH effect\cite{tsui82,laughlin83,hierarchy,jain89,moore91,stormer}, namely the quantized Hall conductance plateaus near various fractional Landau level filling fractions $f$\cite{tsui82}, is not only fundamentally crucial for understanding various types of exotic topological order\cite{wen91}, but is also closely relevant to the technological development of quantum computers\cite{kitaev03,topocompute}.

The FQH plateaus, unlike plateaus near integer fillings whose explanation requires only single-particle physics\cite{laughlin81}, are a consequence of the strong electron-electron interaction combined with (the ever-present) weak disorder in the system. The interaction opens an energy gap to protect the topologically ordered FQH ground state at filling $f$ with a quantized Hall conductance, and the weak disorder localizes excitations to provide a finite Hall conductance plateau near $f$. However, if disorder is much stronger than the interaction scale, the topological FQH ground state and its corresponding plateau will be eventually destroyed, which is consistent with the experimental fact that the FQH effect only exists in samples with high mobility\cite{stormer}. A few numerical studies\cite{dnsheng03,xinwan} of the disorder-driven transition from a topological FQH state to an insulating phase exist, tracking the closing of the energy gap and mobility gap and the collapse of Hall conductance quantization with increasing disorder strength.

Tremendous effort has also gone into understanding FQH physics at fillings where Hall conductance plateaus are absent\cite{west89,zhang92,halperin93,willet93,west14,son15,shao15,jain15,scottscience,sentil16,matteo17,wang17,son17}. A representative example is $f=1/2$, where both experimental\cite{willet93} and theoretical\cite{halperin93,son15,scottscience} work has suggested that composite fermions (CFs)\cite{jain89} form a Fermi sea in clean systems. Although the precise nature of the CF sea is still a central topic of current research\cite{halperin93,son15,scottscience,sentil16,wang17,son17}, early studies already predicted that a transition to an insulator occurs at strong disorder\cite{zhang92,halperin93}.

Following the extensive applications of quantum entanglement spectroscopy methods to clean FQH systems\cite{fqhgammasphere,hli,nicolas09,fqhgammatorus,andreas,ronny10,papic11,PES,zhao12,dubail2012,sterdyniak2012,zaletel2013,wei2015,peterson2015,estienne2015}, corresponding studies in disordered FQH liquids started recently\cite{friedman11,friedman15,zhao2016}. In particular, the ground-state entanglement entropy has been demonstrated as a new and powerful diagnostic of disorder-driven transitions in FQH liquids\cite{zhao2016}. In Ref.~\onlinecite{zhao2016}, we first applied this diagnostic to electrons with Haldane's pseudopotential interaction\cite{hierarchy} at $f=1/3$ in the lowest Landau level (LLL), by tracking the entanglement evolution with increasing disorder. It was shown that the phase transition point can be precisely identified by a sharp increase in the magnitude of the ground-state entanglement entropy derivative with respect to disorder, and a finite-size scaling analysis of the entropy derivative can be used to extract the critical exponent $\nu$ of the diverging correlation length at the transition point. Moreover, it was found that the nearest-neighbor level repulsion statistics of the ground-state entanglement spectrum (for the same system sizes as the entanglement entropy) does not dramatically change at the critical point, and is thus not sensitive to the phase transition.

In this paper, we extend our research of the disorder-driven entanglement evolution to other fractional filling fractions, where the ground states at zero disorder are either incompressible topological FQH states or gapless CF seas. We consider Coulomb interaction between electrons, which is more realistic than the Haldane's pseudopotential used in Ref.~\onlinecite{zhao2016}. The structure of this paper is as follows. We first introduce our model in detail in Sec.~\ref{model}, including the many-body Hamiltonian, the disorder models, the underlying symmetries, and the definitions of the ground-state entanglement entropy. Then in Secs.~\ref{laughlin} and \ref{cf2_5}, we focus on the Laughlin fillings $f=1/3,1/5$ and the hierarchy filling $f=2/5$, where the ground states at zero disorder are incompressible topological FQH states. At all of these fillings, we observe a similar evolution of the ground-state entanglement entropy to that reported in Ref.~\onlinecite{zhao2016}, and the entropy derivative with respect to disorder provides a clear signal for the expected transition from a topological FQH state to an insulator. We extract the length exponent $\nu$ of this transition by a finite-size scaling analysis. In Sec.~\ref{1_2}, we consider $f=1/2$ where the ground state at zero disorder is a gapless CF sea. We examine two disorder models, for both of which we observe a very chaotic behavior of the entropy derivative without a convincing phase transition signal for all probed system sizes within our exact diagonalization (ED) capability. However, the scaling of the entanglement entropy itself versus the system size changes with increasing disorder, which is consistent with the predicted transition from a CF sea to an insulator. In Sec.~\ref{discussion}, we summarize our results, and list some open questions for future work. Finally, we also discuss the entropy evolution at $f=1/7$, and demonstrate the ground-state entanglement spectrum (ES) level statistics at $f=1/3$ for completeness in the Appendix. Similar to the pseudopotential case in Ref.~\onlinecite{zhao2016}, we also observe the localization of the low-energy part of Coulomb ES with increasing disorder, which, however, again only occurs deeply in the insulating phase.

\section{Model}
\label{model}
We consider $N$ interacting electrons in a 2D random potential $U({\bf r})$ on an $L_1\times L_2$ rectangular torus penetrated by a uniform perpendicular magnetic field. We suppose Coulomb interaction $V({\bf r}_i,{\bf r}_j)=\frac{e^2}{\epsilon}\frac{1}{|{\bf r}_i-{\bf r}_j|}$ between electrons, where $-e$ is the electron charge, $\epsilon$ is the dielectric constant, and ${\bf r}_i$ is the coordinate of the $i$th electron. For convenience, we set $e^2/(\epsilon\ell)=1$ and the magnetic length $\ell=1$ as the energy and length units throughout the paper.
In a strong magnetic field, the energy scales of both interaction and disorder are small compared with the Landau level spacing, so we project the many-body Hamiltonian $H=\sum_{i<j}^N V({\bf r}_i-{\bf r}_j)+\sum_{i=1}^N U({\bf r}_i)$ to the LLL, which can be written in the LLL orbital basis as
\begin{eqnarray}
\label{hamil}
H&=&\sum_{m_1,m_2,m_3,m_4=0}^{N_\phi-1}V_{m_1,m_2,m_3,m_4}c_{m_1}^\dagger c_{m_2}^\dagger c_{m_3}c_{m_4}\nonumber\\
&+&\sum_{m_1,m_2=0}^{N_\phi-1}U_{m_1,m_2}c_{m_1}^\dagger c_{m_2}.
\end{eqnarray}
Here $N_\phi=L_1 L_2/(2\pi)$ is the number of magnetic flux quanta penetrating the torus, and $c_m^\dagger$ ($c_m$) creates (annihilates) an electron in the LLL orbital $m$. After choosing the single-particle wave function of orbital $m$ as $\psi_m(x,y)=\Big(\frac{1}{\sqrt{\pi}L_2}\Big)^{\frac{1}{2}}\sum_{n=-\infty}^{+\infty}
e^{\textrm{i}\frac{2\pi}{L_2}(m+nN_\phi)y}e^{-\frac{1}{2}[x-\frac{2\pi}{L_2}(m+nN_\phi)]^2}$, we can compute the interaction matrix elements $V_{\{m_i\}}$ and the disorder matrix elements $U_{\{m_i\}}$ by the standard second-quantization procedure, which gives
\begin{eqnarray}
\label{int}
V_{\{m_i\}}=&&\frac{1}{2}\delta_{m_1+m_2,m_3+m_4}^{{\rm mod} N_\phi}
\sum_{s,t=-\infty}^{+\infty} \delta_{t,m_1-m_4}^{{\rm mod} N_\phi}V_{\bf q}\nonumber\\
&\times&e^{-\frac{1}{2}|{\bf q}|^2}e^{{\rm i}\frac{2\pi s}{N_\phi}(m_1-m_3)}
\end{eqnarray}
and
\begin{eqnarray}
\label{dis}
U_{\{m_i\}}=\sum_{s,t=-\infty}^{+\infty}\delta_{t,m_1-m_2}^{{\rm mod} N_\phi}U_{{\bf q}}
e^{-\frac{1}{4}|{\bf q}|^2}
e^{{\rm i}\frac{\pi s}{N_\phi}(2m_1-t)}.
\end{eqnarray}
Here $\delta^{{\rm mod}N_\phi}_{i,j}$ is the periodic Kronecker delta function with period $N_\phi$, ${\bf q}=(q_x,q_y)=(\frac{2\pi s}{L_1},\frac{2\pi t}{L_2})$ with $|{\bf q}|^2=q_x^2+q_y^2$, $V_{\bf q}=\frac{1}{N_\phi}\frac{1}{|{\bf q}|}$ is the Fourier transform of Coulomb interaction, and $U_{{\bf q}}=\frac{1}{2\pi N_\phi}\int U({\bf r}) e^{-{\rm i}{\bf q}\cdot{\bf r}}d{\bf r}$ is the Fourier transform of $U({\bf r})$. $s=t=0$ must be excluded from the sum in Eq.~(\ref{int}) to remove the artificial divergence (caused by the lack of a positive countercharge in the above model, which is always present in experiments).

We model disorder for the most part using Gaussian white noise, which satisfies $\langle U({\bf r})\rangle=0,\langle U({\bf r})U({\bf r}')\rangle=W^2 \delta({\bf r}-{\bf r}')$ and $\langle U_{\bf q}\rangle=0,\langle U_{\bf q}U_{{\bf q}'}\rangle=\frac{W^2}{2\pi N_\phi} \delta_{{\bf q},-{\bf q}'}$, where $W$ is the strength and $\langle\cdots\rangle$ represents the sample average. In each sample, we generate real $U_{{\bf q=0}}$ from a Gaussian distribution with zero mean and variance $\frac{W^2}{2\pi N_\phi}$. The real part and imaginary part of $U_{{\bf q}\neq{\bf 0}}$ are separately produced from a Gaussian distribution with zero mean and variance $\frac{W^2}{4\pi N_\phi}$. Because $U_{{\bf q}}^*=U_{-{\bf q}}$, the above generation procedures are only implemented for independent $U_{{\bf q}}$'s with ${\bf q}\in\{{\bf q}|q_x=0,q_y\geq 0\}\cup\{{\bf q}|q_x>0\}$. For filling factor $ f = 1/2$, for reasons that will become clear later, we consider, in addition to Gaussian white noise, an ensemble of short-range scatterers corresponding to $U({\bf r})=\sum_n W_n e^{-|{\bf r}-{\bf R}_n|^2/\lambda^2}$ and $U_{{\bf q}}=\frac{\lambda^2}{2N_\phi}\sum_n W_n e^{-\frac{1}{4}\lambda^2 |{\bf q}|^2} e^{-{\rm i} {\bf q}\cdot {\bf R}_n}$. Here $\lambda$ is the range of scatterers, and $W_n$ is the strength of the $n$th scatterer and ${\bf R}_n$ its position, the latter being randomly chosen in each sample.
When averaging over $N_s$ samples, we estimate the error bar of quantity $A$ by $\sqrt{(\langle A^2\rangle-\langle A\rangle ^2)/(N_s-1)}$.

In the absence of disorder, Eq.~(\ref{hamil}) is invariant under the particle-hole (PH) transform $c_m^\dagger\leftrightarrow c_m$ up to a constant shift\cite{Vnote}. A single disorder configuration breaks this symmetry, because the PH transform replaces $U_{\{m_i\}}$ by $-U_{\{m_i\}}^*$ in Eq.~(\ref{hamil}) up to a constant shift, which is equivalent to $U_{q_x,q_y}\rightarrow -U_{q_x,-q_y}$ in Eq.~(\ref{dis}). However, the PH symmetry is statistically preserved by Gaussian white noise, because $-U_{q_x,-q_y}$ still satisfies the Gaussian white noise conditions. Moreover, if the distributions of $W_n$ and $R_n^y$ for an ensemble of scatterers are symmetric with respect to zero, the PH symmetry is also statistically preserved, otherwise it is broken. On the other hand, the magnetic translation invariance conserved in the interaction term (\ref{int}) is always broken by the disorder (\ref{dis}), making the numerical simulation significantly more challenging.

In the following, we choose the isotropic limit with $L_1=L_2=\sqrt{2\pi N_\phi}$. In order to study the entanglement properties, we divide the whole system by two cuts at orbital $m=0$ and $m=\lceil N_\phi/2\rfloor-1$, respectively, where $\lceil x\rfloor$ is the integer part of $x$. This procedure gives two subsystems $A$ and $B$ with boundary length $L=2\sqrt{2\pi N_\phi}$, consisting of orbital $m=0,...,\lceil N_\phi/2\rfloor-1$ and $m=\lceil N_\phi/2\rfloor,...,N_\phi-1$, respectively. We have checked that different positions of the cuts provide the same results statistically for averaged quantities. The entanglement entropy between $A$ and $B$ can be defined as the von-Neumann entropy $S(\rho)=-{\rm Tr}\rho_A\ln\rho_A$, where $\rho_A={\rm Tr}_B\rho$ is the reduced density matrix of part $A$, and $\rho$ is the density matrix describing a suitably chosen ground-state manifold.

For a partially filled LLL at filling $f=N/N_\phi=p/q$ with coprime $p$ and $q$, the ground states of any translation invariant Hamiltonian in clean samples are exactly $D$-fold degenerate with $D\geq q$, guaranteed by the magnetic translation invariance\cite{haldane85}. If $D$ is independent of the system size, such a degeneracy motivates us to consider a ground-state manifold containing the lowest $D$ eigenstates $|\Psi_{i=1,\cdots,D}\rangle$ of the Hamiltonian (\ref{hamil}) at any $W$ for consistency, rather than a single eigenstate. In that case, we will use three choices of $\rho$: \\(i)
\begin{eqnarray}
\overline{\rho}=\frac{1}{D}\sum_{i=1}^D|\Psi_i\rangle\langle\Psi_i|; \nonumber
\end{eqnarray}
(ii)
\begin{eqnarray}
\rho_i=|\Psi_i\rangle\langle\Psi_i|; \nonumber
\end{eqnarray}
and (iii)
\begin{eqnarray}
\rho_i^{\min}=|\Psi_{\min}^i\rangle\langle\Psi_{\min}^i|, \nonumber
\end{eqnarray}
where $|\Psi_{\min}^i\rangle$ is the minimally entangled state (MES)\cite{zhangyi} in the ground-state manifold.

Depending on different choices, we correspondingly measure the entanglement entropy by $S(\overline{\rho})$, $\overline{S}=\frac{1}{D}\sum_{i=1}^D S(\rho_i)$, and $\overline{S}_{\min}=\frac{1}{M}\sum_{i=1}^M S(\rho_i^{\min})$ respectively\cite{zhao2016}, where $M$ is the number of MESs. The sum over states in $\overline{S}$ and $\overline{S}_{\min}$ is to minimize the effect of statistical fluctuations for a finite number of samples of finite size. However, if $D$ depends on the system size, we just choose a ground-state manifold only containing the lowest eigenstate $|\Psi_1\rangle$ of the Hamiltonian (\ref{hamil}), i.e. $\rho=|\Psi_1\rangle\langle\Psi_1|$, so the ground-state entanglement entropy is measured by $S(\rho)=S(|\Psi_1\rangle)$.

\section{Laughlin fillings}
\label{laughlin}
We first consider fillings $f=1/q$ with $q=3$ and $5$. In clean systems, we have numerically confirmed for various system sizes that the Coulomb ground states at these fillings are always exactly $q$-fold degenerate, so we choose the lowest $q$ eigenstates of the Hamiltonian (\ref{hamil}) as the ground-state manifold. The overlap and energy gap calculations at these fillings suggest that the Coulomb ground states in clean systems are gapped and well captured by the Laughlin model states, although the deviation from the model states increases for larger $q$ (Table~\ref{overlap}). In the non-interacting limit with $W=\infty$ for PH-symmetric disorder, because extended single-particle states only exist at the center of the LLL band\cite{huo}, all occupied single-particle states below the Fermi level at $f=1/3$ and $1/5$ are localized, which means that the ground state is an Anderson insulator. Therefore, we expect a transition from the topological Laughlin phase to an insulating phase with increasing disorder at these fillings. In the following, we will characterize this transition by the entanglement entropy of the ground-state manifold, with disorder modeled by Gaussian white noise.

\subsection{$f=1/3$}
For $q=3$, the Coulomb ground states at zero disorder are very well described by the $f=1/3$ Laughlin model states, as indicated by the extremely high overlaps that are stable against increasing system size (Table~\ref{overlap}). The Hall conductance plateau at $f=1/3$ is the first reported FQH effect in experiments\cite{tsui82}. Strong disorder closes the energy gap and the mobility gap, leading to a phase transition to an insulating phase\cite{dnsheng03,xinwan}. We compute the ground-state manifold by ED for $N\leq 10$ electrons with Hilbert space dimension up to $30045015$, then monitor the evolution of its entanglement entropy with increasing disorder.

\begin{table}
\caption{The squared overlap $\mathcal{O}$ at zero disorder between the Coulomb ground state and the corresponding Laughlin model state, and the energy difference $\Delta$ between the $q$th and the $(q+1)$th lowest eigenstate of the Hamiltonian (\ref{hamil}) at zero disorder, at $f=1/3$ and $1/5$. Laughlin model states are obtained by diagonalizing their parent Hamiltonians, i.e., Haldane's pseudopotentials. In topological FQH phases, while the overlap is expected to decrease as the system size gets very large, the energy gap is expected to remain robust, as can be seen for $f = 1/3$ and $1/5$. We also provide the data at $f=1/7$, which will be used in Appendix~\ref{appa}.}
\begin{ruledtabular}
\begin{tabular}{ccccccc}
&\multicolumn{2}{c}{$f=1/3$}&\multicolumn{2}{c}{$f=1/5$}&\multicolumn{2}{c}{$f=1/7$}\\
\hline
&$\mathcal{O}$&$\Delta$&$\mathcal{O}$&$\Delta$&$\mathcal{O}$&$\Delta$\\
$N=4$&$0.9788$&$0.0472$&$0.9528$&$0.0080$&$0.9555$&$0.0028$\\
$N=5$&$0.9947$&$0.0631$&$0.9556$&$0.0101$&$0.8613$&$0.0028$\\
$N=6$&$0.9891$&$0.0630$&$0.9222$&$0.0107$&$0.8575$&$0.0033$\\
$N=7$&$0.9921$&$0.0624$&$0.9351$&$0.0091$&$0.7997$&$0.0022$\\
$N=8$&$0.9803$&$0.0603$&$0.7932$&$0.0062$&$0.5230$&$0.0006$\\
$N=9$&$0.9694$&$0.0585$&$0.7199$&$0.0070$& & \\
$N=10$&$0.9755$&$0.0630$&$0.8636$&$0.0110$& & \\
$N=11$&$0.9768$&$0.0651$& & & & \\
$N=12$&$0.9744$&$0.0631$& & & & \\
$N=13$&$0.9725$&$0.0629$& & & &
\end{tabular}
\end{ruledtabular}
\label{overlap}
\end{table}

\begin{figure*}
\centerline{\includegraphics[width=\linewidth]{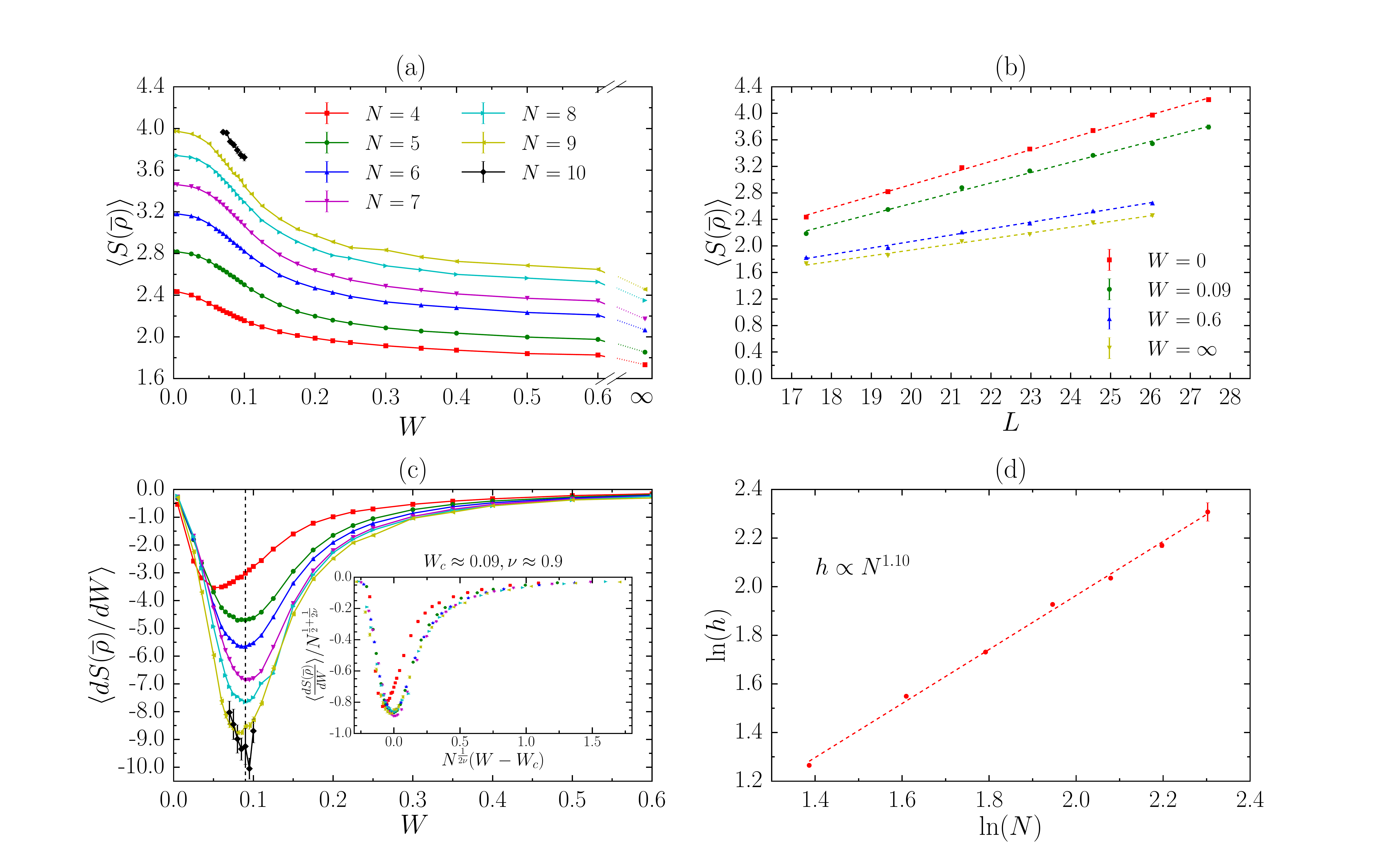}}
\caption{Entanglement entropy measured by $S(\overline{\rho})$ for $N=4-10$ electrons at $f=1/3$. (a)
$\langle S(\overline{\rho}) \rangle$ versus $W$. (b) $\langle S(\overline{\rho}) \rangle$ versus the cut length $L$ at $W=0,0.09,0.6$, and $\infty$. The dashed line is the linear fitting of $\langle S(\overline{\rho}) \rangle$ versus $L$. (c) $\langle dS(\overline{\rho})/dW\rangle$ versus
$W$, replotted in terms of scaled variables $\langle dS(\overline{\rho})/dW\rangle/N^{\frac{1}{2}+\frac{1}{2\nu}}$ and $N^{\frac{1}{2\nu}}(W-W_c)$ in the inset with $W_c\approx 0.09$ and $\nu\approx 0.9$. The vertical dashed line indicates $W=0.09$. Each color corresponds to the same system size as in (a). (d) The minimum magnitude $h$ versus $N$ on a double logarithmic scale. The dashed line corresponds to $h\propto N^{1.10}$. We averaged $20000$ samples for $N=4-7$, $5000$ samples for $N=8$, $800$ samples for $N=9$, and $50$ samples for $N=10$ electrons. The calculation for $N=10$ is only done at a few points near $W=0.09$. The results are consistent with those for smaller systems sizes, but with much larger error bars due to much fewer samples. In (a), we also give the data at $W=\infty$.}
\label{S_1_3}
\end{figure*}

\begin{figure*}
\centerline{\includegraphics[width=\linewidth]{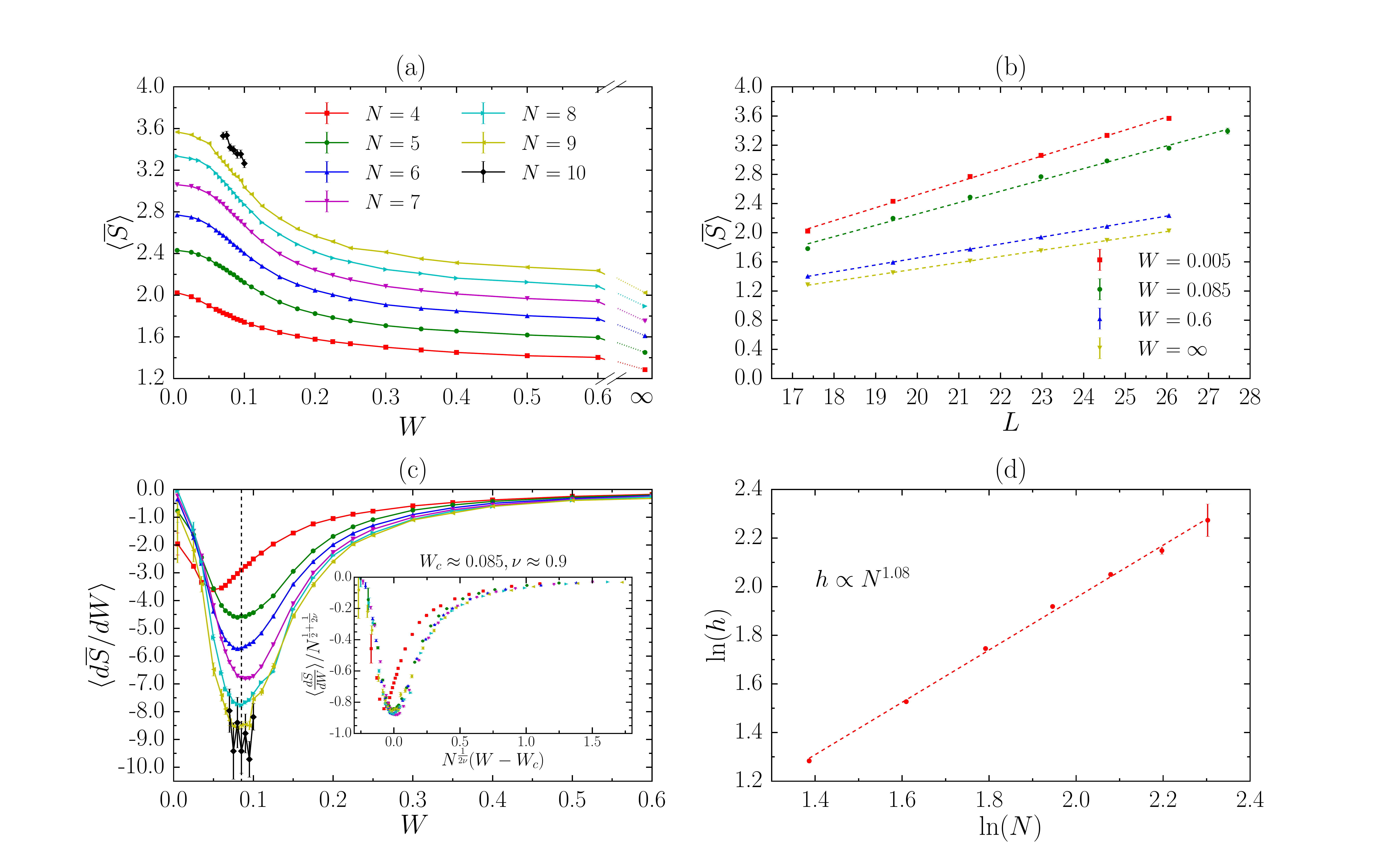}}
\caption{Entanglement entropy measured by $\overline{S}$ for $N=4-10$ electrons at $f=1/3$. (a)
$\langle \overline{S} \rangle$ versus $W$. (b) $\langle \overline{S} \rangle$ versus the cut length $L$ at $W=0,0.085,0.6$, and $\infty$. The dashed line is the linear fitting of $\langle \overline{S} \rangle$ versus $L$. (c) $\langle d\overline{S}/dW\rangle$ versus
$W$, replotted in terms of scaled variables $\langle d\overline{S}/dW\rangle/N^{\frac{1}{2}+\frac{1}{2\nu}}$ and $N^{\frac{1}{2\nu}}(W-W_c)$ in the inset with $W_c\approx 0.085$ and $\nu\approx 0.9$. The vertical dashed line indicates $W=0.085$. Each color corresponds to the same system size as in (a). (d) The minimum magnitude $h$ versus $N$ on a double logarithmic scale. The dashed line corresponds to $h\propto N^{1.10}$. We averaged $20000$ samples for $N=4-7$, $5000$ samples for $N=8$, $800$ samples for $N=9$, and $50$ samples for $N=10$ electrons. The calculation for $N=10$ is only done at a few points near $W=0.085$. The results are consistent with those for smaller systems sizes, but with much larger error bars due to much fewer samples. In (a), we also give the data at $W=\infty$.}
\label{S_1_3psi}
\end{figure*}

We first measure the entanglement entropy by $S(\overline{\rho})$. Similar to the case of $f=1/3$ with Haldane's pseudopotential\cite{zhao2016}, we find that $\langle S(\overline{\rho}) \rangle$ decreases with $W$ for a fixed system size [Fig.~\ref{S_1_3}(a)]. However, it increases with the system size at a fixed $W$, always agreeing with an area law $\langle S(\overline{\rho}) \rangle\propto L$ [Fig.~\ref{S_1_3}(b)]. We further compute the derivative of $S(\overline{\rho})$ with respect to the disorder strength, $dS(\overline{\rho})/dW$, approximated in each sample by $[S(\overline{\rho})|_{W+\Delta W}-S(\overline{\rho})|_W]/\Delta W$ with $\Delta W=0.001W$, where only the magnitude of $W$ is changed by a small percentage but the disorder configuration is kept fixed. One can see that all $\langle dS(\overline{\rho})/dW\rangle$ curves exhibit a pronounced minimum that gets deeper for larger systems [Fig.~\ref{S_1_3}(c)]. Except the smallest $N=4$, this minimum is located at $W_c\approx 0.09$, which is almost independent of the system size. As demonstrated in a double logarithmic plot [Fig.~\ref{S_1_3}(d)], the magnitude of the minimum $h=|\min_W\langle dS(\overline{\rho})/dW\rangle|$ grows with $N$ as $h\propto N^{1.10}$, which is consistent with a divergence in the thermodynamic limit. Informed by the fact that thermal phase transitions are very often characterized by a singularity in the specific heat (which is proportional to the temperature derivative of thermal entropy), we consider this divergence of the disorder derivative of entanglement entropy as a convincing signature of the expected quantum phase transition from the $f=1/3$ Laughlin phase to an insulating phase. A sharp drop in the entanglement entropy and a pronounced peak in the entanglement derivative were also used to identify a first-order transition in clean bilayer quantum Hall systems as a function of layer separation\cite{john11}.

For a continuous phase transition in our case, the area law shown in Fig.~\ref{S_1_3}(b) suggests a scaling behavior
\begin{eqnarray}
S(\overline{\rho}) \propto N^{\frac{1}{2}}f[N^{\frac{1}{2\nu}}(W-W_c)],\nonumber
\end{eqnarray}
for large $N$ (here we have used $L=2\sqrt{2\pi N_\phi}=2\sqrt{2\pi N/f}\propto \sqrt{N}$), leading to
\begin{eqnarray}
dS(\overline{\rho})/dW\propto N^{\frac{1}{2}+\frac{1}{2\nu}}f'[N^{\frac{1}{2\nu}}(W-W_c)]\nonumber
\end{eqnarray}
and, in particular,
\begin{eqnarray}
dS(\overline{\rho})/dW|_{W=W_c}\propto N^{\frac{1}{2}+\frac{1}{2\nu}},\nonumber
\end{eqnarray}
where $f'$ means derivative. Thus $h\propto N^{1.10}$ in Fig.~\ref{S_1_3}(d) implies that $\nu\approx 0.9$. By plotting the rescaled variable $\langle dS(\overline{\rho})/dW\rangle/N^{\frac{1}{2}+\frac{1}{2\nu}}$ versus $N^{\frac{1}{2\nu}}(W-W_c)$ for $W_c\approx 0.09$ and $\nu\approx 0.9$, we indeed find that besides the smallest size $N=4$, all data in Fig.~\ref{S_1_3}(c) collapse onto a single curve [Fig.~\ref{S_1_3}(c) inset].

$W_c$ obtained from Coulomb interaction is very different from its counterpart obtained from Haldane's pseudopotential\cite{zhao2016}. For Coulomb interaction, we find $W_c\approx 0.09$, which is seven times smaller than the reported value $\approx 0.6$ for Haldane's pseudopotential, reflecting the fact that Coulomb ground states are protected by a smaller gap and are more fragile against disorder. Since $\nu$ is a critical exponent, however, we would expect that Coulomb interaction and pseudopotential interaction should give roughly the same $\nu$. However, we numerically get $\nu\approx 0.9$ for Coulomb interaction, which is $50\%$ larger than the reported value $\approx 0.6$ for Haldane's pseudopotential and almost reaches the conventional $\nu\geq 2/d$ bound for $d-$dimensional disordered systems\cite{harris,Chayes86}. We have further examined an interpolation between Coulomb and Haldane's pseudopotential, and observed a continuous varying of $\nu$. Such an apparent dependence of $\nu$ on the interaction suggests that corrections to finite-size scaling are still significant in the system sizes reached by ED.

An alternative measure of the entanglement entropy is given by $\overline{S}$. Because the ground states in clean systems are exactly degenerate and different choices of ground states may lead to very different $\overline{S}$\cite{zhangyi}, this quantity is not well defined at $W=0$. This also causes the larger error bars of $\langle d\overline{S}/dW\rangle$ compared to $\langle dS(\overline{\rho})/dW\rangle$ at very small $W$. However, once the disorder is not too weak, the results of $\overline{S}$ (Fig.~\ref{S_1_3psi}) are very similar to those of $S(\overline{\rho})$. Remarkably, the minimum of $\langle d\overline{S}/dW\rangle$ is located at $W_c\approx 0.085$, which is almost the same as that of $\langle dS(\overline{\rho})/dW\rangle$ [Fig.~\ref{S_1_3psi}(c)]. The finite-size scaling analysis also gives a similar $\nu\approx 0.9$ [Fig.~\ref{S_1_3psi}(c) inset].

\begin{figure}
\centerline{\includegraphics[width=\linewidth]{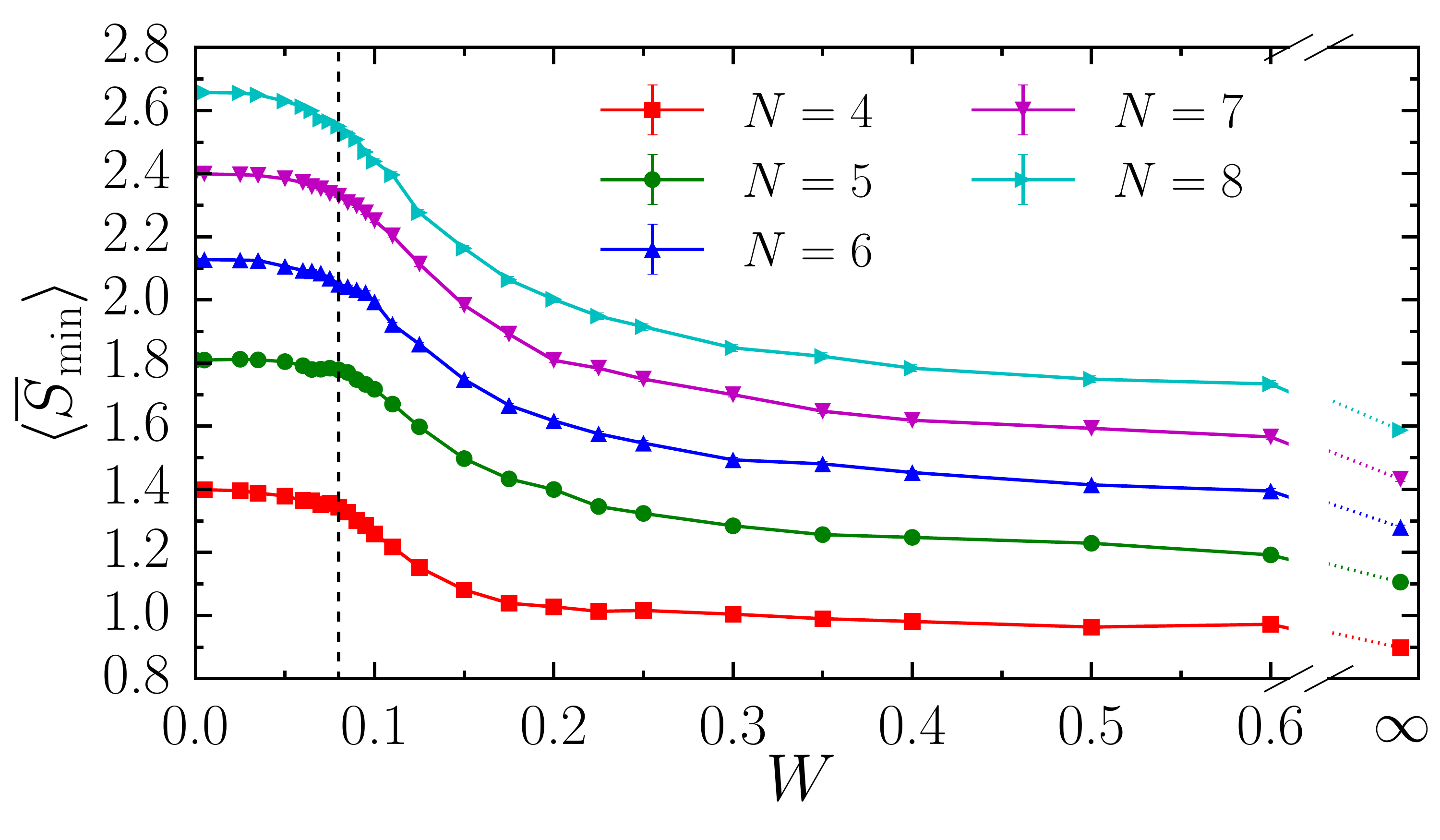}}
\caption{Entanglement entropy measured by $\overline{S}_{\min}$ versus
$W$ for $N=4-8$ electrons at $f=1/3$. We averaged $2000$ samples for $N=4-7$ and $1000$ samples for $N=8$ electrons. The vertical dashed line indicates $W=0.08$. The data at $W=\infty$ are also given.}
\label{mes_1_3c}
\end{figure}

Finally, we study the entanglement entropy of the MES\cite{zhangyi,wei1,wei2} in the ground-state manifold. We consider all superpositions $|\Psi\rangle=\sin\theta_1\sin\theta_2|\Psi_1\rangle+\sin\theta_1\cos\theta_2 e^{{\rm i}\phi_1}|\Psi_2\rangle+\cos\theta_1 e^{{\rm i}\phi_2}|\Psi_3\rangle$ with $\theta_1,\theta_2\in[0,\pi/2]$ and $\phi_1,\phi_2\in[0,2\pi)$, then numerically search for the local minima of $S(|\Psi\rangle)$ in the parameter space spanned by $(\theta_1,\theta_2,\phi_1,\phi_2)$. $|\Psi\rangle$'s at these local minima correspond to the MES $|\Psi_{\min}^{i=1,\cdots,M}\rangle$. Since searching for these MESs is a complicated four-dimensional minimization problem, we can only reach $N=8$ electrons with less samples, and do not perform the calculation of $\langle d\overline{S}_{\min}/dW\rangle$. Instead, in Fig.~\ref{mes_1_3c}, we show $\langle \overline{S}_{\min}\rangle$ as a function of $W$. At small disorder, an almost constant $\langle \bar{S}_{\min}\rangle$ suggests that the ground-state topological properties are the same as those in clean systems. However, the plateau of $\langle \bar{S}_{\min}\rangle$ is not as good as that for Haldane's pseudopotential\cite{zhao2016}. We attribute this as being due to the larger finite-size effect in Coulomb ground states. $\langle \bar{S}_{\min}\rangle$ starts to significantly drop at $W\approx0.08$, signifying a transition point consistent with those indicated by $\langle dS(\overline{\rho})/dW\rangle$ and $\langle d\overline{S}/dW\rangle$.

\begin{figure*}
\centerline{\includegraphics[width=\linewidth]{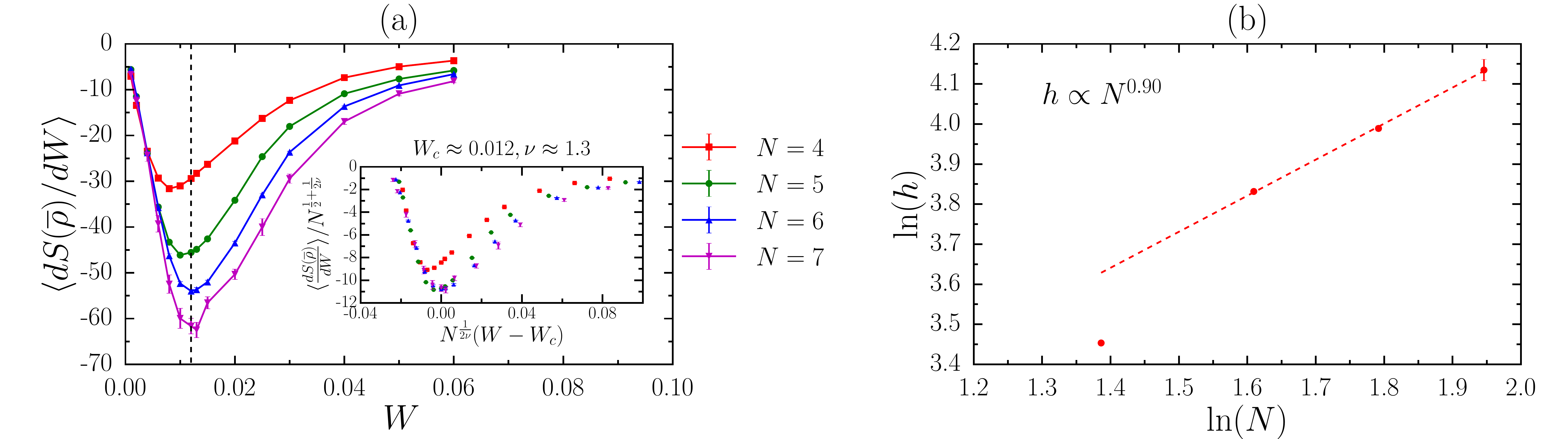}}
\caption{Entanglement entropy measured by $S(\overline{\rho})$ for $N=4-7$ electrons at $f=1/5$. (a) $\langle dS(\overline{\rho})/dW\rangle$ versus
$W$, replotted in terms of scaled variables $\langle dS(\overline{\rho})/dW\rangle/N^{\frac{1}{2}+\frac{1}{2\nu}}$ and $N^{\frac{1}{2\nu}}(W-W_c)$ in the inset with $W_c\approx 0.012$ and $\nu\approx 1.3$. The vertical dashed line indicates $W=0.012$. (b) The depth of the minimum $h$ versus $N$ on a double logarithmic scale. The dashed line corresponds to $h\propto N^{0.90}$. We averaged $20000$ samples for $N=4$ and $5$, $2000$ samples for $N=6$, and $100$ samples for $N=7$ electrons.}
\label{dsdw_1_5}
\end{figure*}

In summary, at $f=1/3$, all of the three entanglement measurements give consistent identifications of the transition from the Laughlin phase to an insulating phase. We have examined that this consistency also holds for $f=1/5$ and $2/5$. Therefore, we will only demonstrate the results of $dS(\overline{\rho})/dW$ in the remainder of this section as well as in the following section.

\subsection{$f=1/5$}
For $q=5$, the overlaps between the ground states at zero disorder and the $f=1/5$ Laughlin model states are still high (Table~\ref{overlap}), which is consistent with the experimental observation of a robust FQH effect in high-quality samples at $f=1/5$\cite{jiang}. However, compared with the $f=1/3$ case, the deviation from the model states is larger, and the energy gaps are smaller, implying that the topological ground states at $f=1/5$ are more fragile against disorder than those at $f=1/3$. In the following, we compute the ground-state manifold by ED for $N\leq 7$ electrons with Hilbert space dimension up to $6724520$, and track their entanglement entropy evolution. The reason why we can only reach a smaller $N$ at $f=1/5$ than at $f=1/3$ is that the Hilbert space for a fixed $N$ is larger at lower fillings.

We show $\langle dS(\overline{\rho})/dW\rangle$ in Fig.~\ref{dsdw_1_5}. Similar to the $f=1/3$ case, we observe a pronounced minimum in all $\langle dS(\overline{\rho})/dW\rangle$ curves at $f=1/5$, whose magnitude increases with the system size [Fig.~\ref{dsdw_1_5}(a)]. However, as expected from the overlap and energy gap calculations at zero disorder, this minimum is located at a much smaller $W_c\approx 0.012$ (except for the smallest system size $N=4$) than that $\approx 0.09$ at $f=1/3$. The minimum magnitude $h$ also shows a larger finite-size effect than the $f=1/3$ case. At $f=1/5$, the data point of $N=4$ in the $\ln h-\ln N$ plot obviously deviates from the linear growth of other three points [Fig.~\ref{dsdw_1_5}(b)]. With the point of $N=4$ neglected, we obtain $h\propto N^{0.90}$, which suggests $\nu\approx 1.3$ according to the finite-size scaling. Indeed, if we set $W_c\approx 0.012$ and $\nu\approx 1.3$, all data in Fig.~\ref{dsdw_1_5}(a) collapse to a single rescaled curve for $N=5-7$ electrons [Fig.~\ref{dsdw_1_5}(a) inset]. We notice that $\nu$ obtained at $f=1/5$ is larger than that $\approx 0.9$ at $f=1/3$, and is consistent with the $\nu\geq 2/d$ bound. 

\begin{figure*}
\centerline{\includegraphics[width=\linewidth]{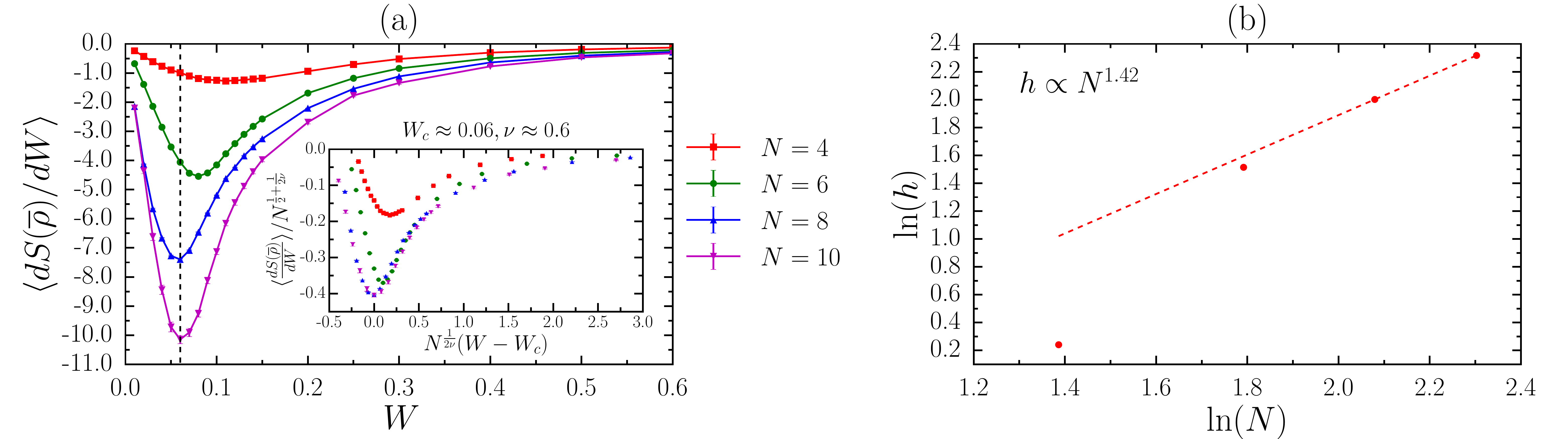}}
\caption{Entanglement entropy measured by $S(\overline{\rho})$ for $N=4,6,8$, and $10$ electrons at $f=2/5$. (a) $\langle dS(\overline{\rho})/dW\rangle$ versus
$W$, replotted in terms of scaled variables $\langle dS(\overline{\rho})/dW\rangle/N^{\frac{1}{2}+\frac{1}{2\nu}}$ and $N^{\frac{1}{2\nu}}(W-W_c)$ in the inset with $W_c\approx 0.06$ and $\nu\approx 0.6$. The vertical dashed line indicates $W=0.06$. (b) The depth of the minimum $h$ versus $N$ on a double logarithmic scale. The dashed line corresponds to $h\propto N^{1.42}$. We averaged $20000$ samples for $N=4,6,8$, and $400$ samples for $N=10$ electrons.}
\label{dsdw_2_5}
\end{figure*}

\section{$f=2/5$}
\label{cf2_5}
We next go beyond the Laughlin fillings and consider $f=2/5$. A robust FQH effect was experimentally observed at this filling\cite{stormer83}, which can be interpreted as two fully filled effective Landau levels of CFs\cite{jain89}, or the daughter of the $f=1/3$ FQH effect in the hierarchy scenario\cite{hierarchy}. At zero disorder, the Coulomb ground states at $f=2/5$ are incompressible FQH states with large overlaps with the CF ansatz wave functions\cite{Hermanns,hermann13}. Here again, for PH symmetric disorder, in the non-interacting limit, all single-particle states below the Fermi level at $f=2/5$ are localized, leading to an Anderson insulator. Therefore, we again expect a transition from a topological FQH phase to an insulator with increasing disorder at this filling. Since the Coulomb ground states at $f=2/5$ are fivefold degenerate at zero disorder for all system sizes, we choose the ground-state manifold containing the lowest five eigenstates of the Hamiltonian (\ref{hamil}), which is obtained by ED for $N\leq 10$ electrons with Hilbert space dimension up to $3268760$. We model disorder by Gaussian white noise in this section.

The evolution of entanglement entropy at $f=2/5$ is similar to those at Laughlin fillings (Fig.~\ref{dsdw_2_5}). We observe a pronounced minimum in all $\langle dS(\overline{\rho})/dW\rangle$ curves. The position of this minimum stays around $W_c\approx 0.06$ for $N\geq 8$ electrons [Fig.~\ref{dsdw_2_5}(a)], which means that the FQH phase with Coulomb interaction at $f=2/5$ is more robust against disorder than that at $f=1/5$. There is a large finite-size effect in the minimum magnitude $h$: the data points of $N=4$ and $N=6$ electrons in the $\ln h-\ln N$ plot deviate from the linear growth of other two points [Fig.~\ref{dsdw_2_5}(b)]. Based on the finite-size scaling analysis of the data for $N=8$ and $N=10$ electrons, we obtain $\nu\approx 0.6$ [Fig.~\ref{dsdw_2_5}(a) inset]. This value is smaller than those at $f=1/3$ and $f=1/5$, and again violates the $\nu\geq 2/d$ bound.

\section{$f=1/2$}
\label{1_2}
Having considered several filling fractions where incompressible FQH phases exist in clean systems, we now extend our discussion to $f=1/2$. For the half-filled Landau level without disorder, in a mean-field picture CFs feel a zero effective magnetic field and consequently form a gapless CF sea instead of an incompressible FQH state\cite{halperin93,willet93,son15,scottscience,matteo17}. For finite systems, the shape of such a CF sea depends on the system size, leading to a variable ground-state degeneracy $D$, as shown in Table~\ref{1_2_GSD}. Because the averaging method used in Secs.~\ref{laughlin} and \ref{cf2_5} is not appropriate for a manifold with a degeneracy depending on the system size, here we just focus on the entanglement entropy of the lowest eigenstate $|\Psi_1\rangle$ of the Hamiltonian (\ref{hamil}), which is obtained by ED for $N\leq 13$ electrons with Hilbert space dimension up to $10400600$.

\begin{table}
\caption{Ground-state degeneracy $D$ in clean systems for $N=6-13$ electrons at $f=1/2$.}
\begin{ruledtabular}
\begin{tabular}{ccccccccc}
$N$&$6$&$7$&$8$&$9$&$10$&$11$&$12$&$13$\\
\hline
$D$&$4$&$8$&$8$&$2$&$8$&$16$&$2$&$8$
\end{tabular}
\end{ruledtabular}
\label{1_2_GSD}
\end{table}

Naively one may expect that the metallic CF sea in clean systems will be destroyed by an arbitrarily small disorder, according to the scaling theory of localization, which excludes the metallic behavior in 2D non-interacting systems with random disorder if the magnetic field is absent\cite{anderson}. However, being different from ordinary fermions, CFs carry magnetic flux. Early arguments suggested\cite{zhang92} that the disorder-induced inhomogeneous CF density produces random local fluctuations of the effective magnetic field (although the net effective magnetic field is still zero), thus suppressing the localization of CFs and stabilizing the metallic CF sea at weak disorder\cite{Comment}. Then, at strong disorder, the transition to an insulator occurs, consistent with the conventional localization theory. In the following, we will look for the clue of this phase transition in the entanglement entropy.

\begin{figure*}
\centerline{\includegraphics[width=\linewidth]{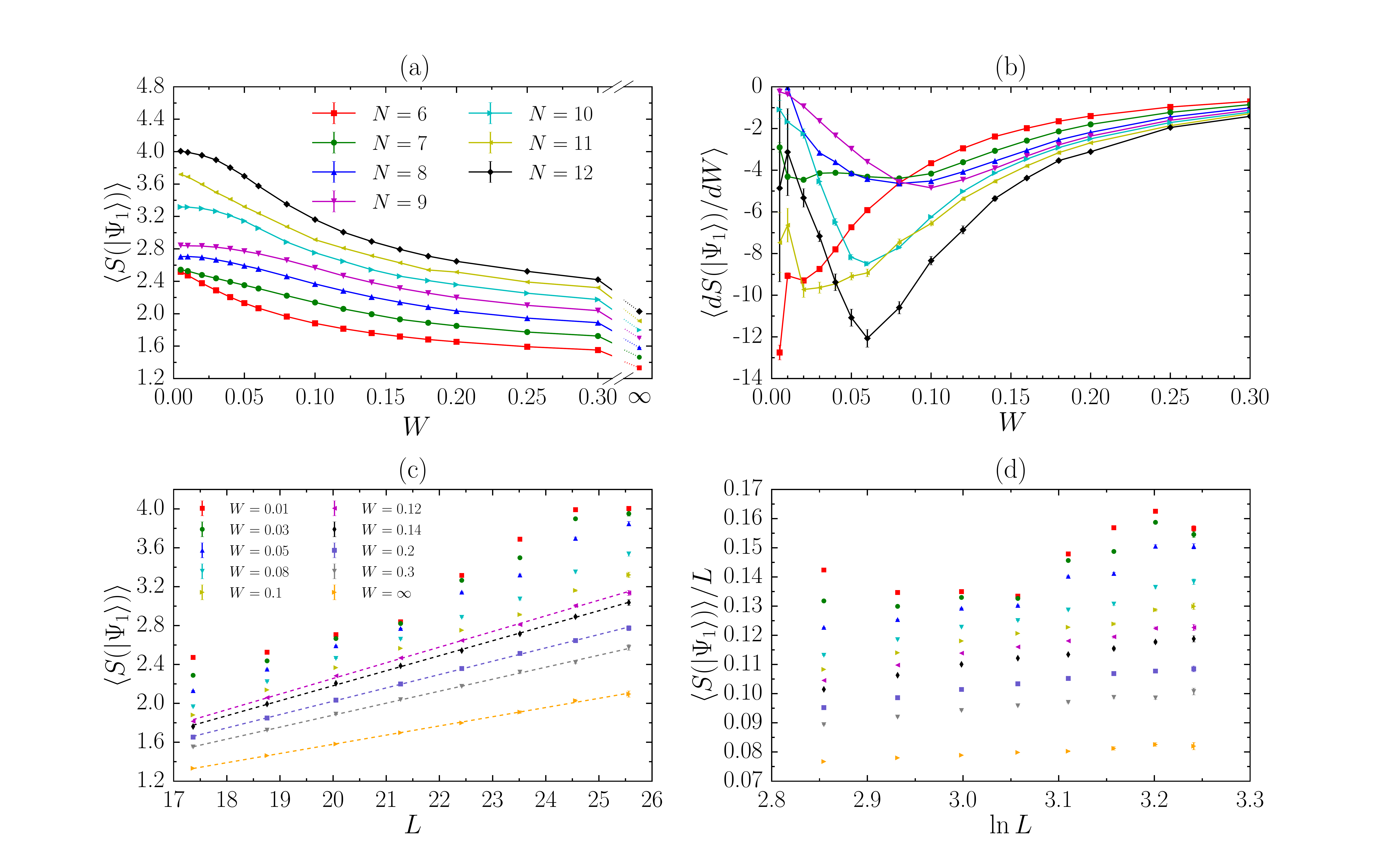}}
\caption{Entanglement entropy measured by $S(|\Psi_1\rangle)$ at $f=1/2$ with Gaussian white noise. (a) $\langle S(|\Psi_1\rangle)\rangle$ versus $W$ for $N=6-12$ electrons. (b) $\langle dS(|\Psi_1\rangle)/dW\rangle$ versus $W$ for $N=6-12$ electrons. Each color corresponds to the same system size as in (a). (c) $\langle S(|\Psi_1\rangle)\rangle$ versus $L$ at various $W$ for $N=6-13$ electrons. The dashed line is the linear fitting of $\langle S(|\Psi_1\rangle)\rangle$ versus $L$. (d) $\langle S(|\Psi_1\rangle)\rangle /L$ versus $\ln L$ at various $W$ for $N=6-13$ electrons. Each color corresponds to the same $W$ as in (c). We averaged $20000$ samples for $N=6-9$, $5000$ samples for $N=10$, $500$ samples for $N=12$, and $100$ samples for $N=13$ electrons. In (a), (c), and (d), we also give the data at $W=\infty$.}
\label{GWN_1_2}
\end{figure*}

\begin{figure*}
\centerline{\includegraphics[width=\linewidth]{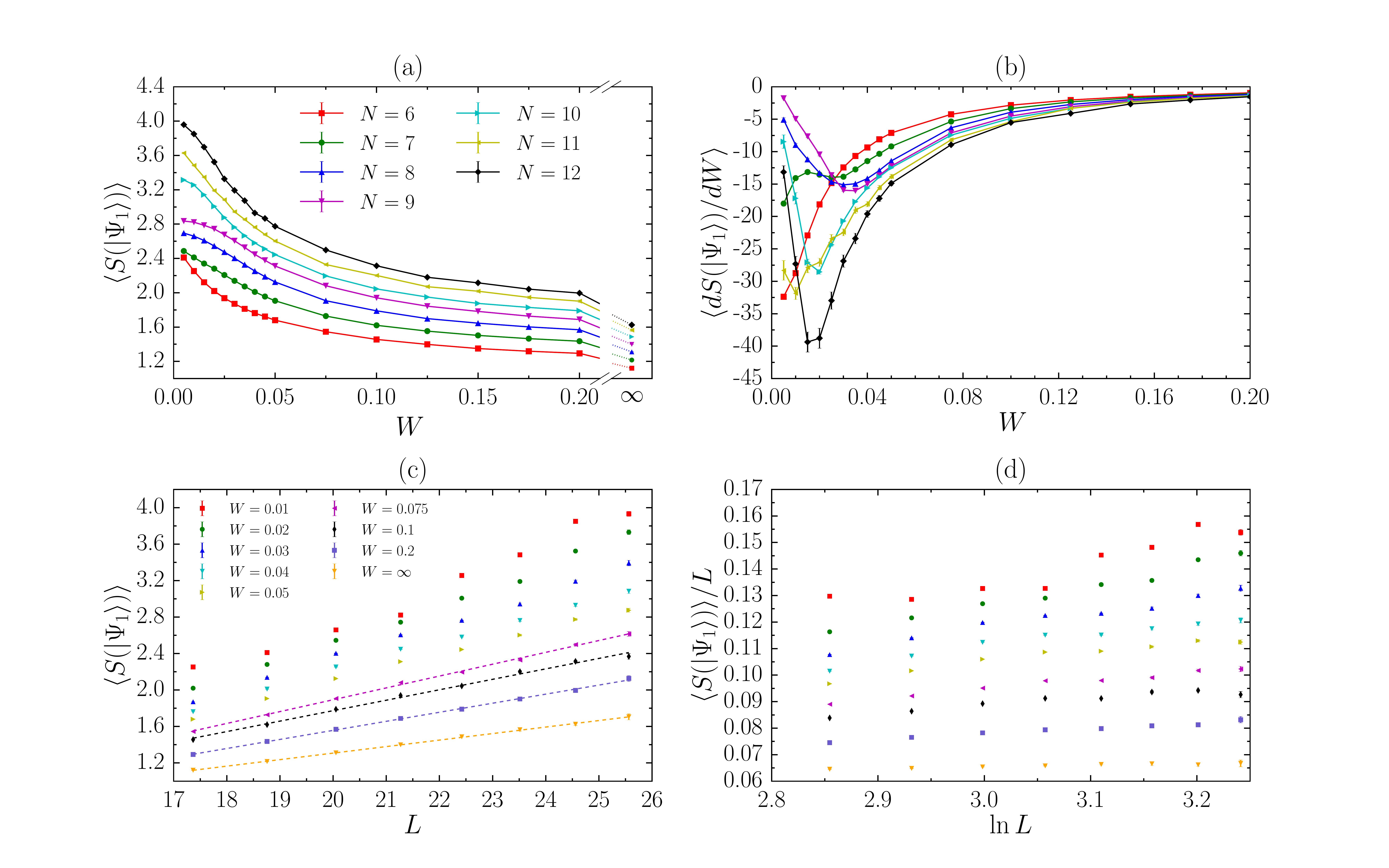}}
\caption{Entanglement entropy measured by $S(|\Psi_1\rangle)$ at $f=1/2$ with an ensemble of scatterers. (a) $\langle S(|\Psi_1\rangle)\rangle$ versus $W$ for $N=6-12$ electrons. (b) $\langle dS(|\Psi_1\rangle)/dW\rangle$ versus $W$ for $N=6-12$ electrons. Each color corresponds to the same system size as in (a). (c) $\langle S(|\Psi_1\rangle)\rangle$ versus $L$ at various $W$ for $N=6-13$ electrons. The dashed line is the linear fitting of $\langle S(|\Psi_1\rangle)\rangle$ versus $L$. (d) $\langle S(|\Psi_1\rangle)\rangle /L$ versus $\ln L$ at various $W$ for $N=6-13$ electrons. Each color corresponds to the same $W$ as in (c). We averaged $20000$ samples for $N=6-9$, $5000$ samples for $N=10$, $500$ samples for $N=12$, and $100$ samples for $N=13$ electrons. In (a), (c), and (d), we also give the data at $W=\infty$.}
\label{scatter_1_2}
\end{figure*}

We first consider Gaussian white noise, which preserves the PH symmetry. $S(|\Psi_1\rangle)$ and $\langle dS(|\Psi_1\rangle)/dW\rangle$ are shown in Figs.~\ref{GWN_1_2}(a) and \ref{GWN_1_2}(b), respectively. Strikingly, we observe a chaotic behavior of the entropy derivative at $f=1/2$, which is very different from those at $f=1/3,1/5$, and $2/5$. For some system sizes such as $N=6$ and $N=7$ electrons, it is difficult to identify a minimum in $\langle dS(|\Psi_1\rangle)/dW\rangle$. For other system sizes where an obvious minimum in $\langle dS(|\Psi_1\rangle)/dW\rangle$ exists, its position is still changing significantly with the system size, and the magnitude does not nicely scale with $N$. Therefore, at least for the system sizes that we can reach by ED, the disorder derivative of the entanglement entropy with Gaussian white noise does not provide a convincing signal of the phase transition at $f=1/2$.

As a result of the PH symmetry, the $f=1/2$ ground state in the $W=\infty$ limit for Gaussian white noise is not an insulator, but a metallic critical phase with the same Hall and longitudinal conductance $\sigma_{xy}=\sigma_{xx}=0.5e^2/h$\cite{huo93}, which may make Gaussian white noise inappropriate for the study of a transition at $f=1/2$ from a CF sea to an insulator. In order to understand whether the chaotic $\langle dS(|\Psi_1\rangle)/dW\rangle$ observed above is due to the absence of an insulator in the non-interacting limit, we then consider a different disorder model that breaks the PH symmetry. We choose an ensemble of scatterers\cite{huo93}, and assume that (i) the range of scatterers is $\lambda=1/\sqrt{2}$, which is comparable with the magnetic length $\ell=1$; (ii) the number of scatterers in the ensemble is $3N_\phi$, which is significantly more than one per flux quantum; (iii) $W$ is negative, and $W_n=10W$ for $20\%$ of the scatterers and $W_n=W$ for the remaining in each sample; (iv) the distribution of ${\bf R}_n$ is symmetric with respect to zero. In the non-interacting limit, these settings significantly skew the density of states and shift the position of extended single-particle states from LLL filling $f_c=1/2$ to $f_c\approx 0.6$, thus breaking the PH symmetry. The $f=1/2$ ground state at $W=\infty$ for such an ensemble of scatterers is hence an Anderson insulator with $\sigma_{xy}=\sigma_{xx}=0$. However, even in this case, we still observe a very chaotic $\langle dS(|\Psi_1\rangle)/dW\rangle$ [Fig.~\ref{scatter_1_2}(b)]. $S(|\Psi_1\rangle)$ [Fig.~\ref{scatter_1_2}(a)] is also similar to that for Gaussian white noise. Therefore, our numerical data suggest that, for the system sizes that we can reach by ED, the significant size-dependence in the disorder derivative of ground-state entanglement entropy at $f=1/2$ is not significantly affected by the choice of the disorder model, or the state in the infinite disorder limit, but is more likely due to the absence of a gap at zero disorder.

It is known that a logarithmic correction to the area law of entanglement entropy is expected if a Fermi surface is present, like in the case of CF sea\cite{wolf06,shao15}. However, if disorder really induces the collapse of the CF sea, the area law should be recovered at strong disorder. Therefore we further examine the scaling of $S(|\Psi_1\rangle)$ versus the boundary length $L$ of the subsystem $A$ at various disorder. For both of the Gaussian white noise and the scatterer ensemble, we indeed find that $S(|\Psi_1\rangle)$ grows with $L$ faster than a linear scaling at small disorder, then gradually evolves to a linear scaling with increasing disorder [Figs.~\ref{GWN_1_2}(c) and \ref{scatter_1_2}(c)]. This tendency is consistent with the predicted transition from a CF sea to an insulator. Although our system sizes are still too small to manifest $S/L\propto \ln L$ at small disorder [Figs.~\ref{GWN_1_2}(d) and \ref{scatter_1_2}(d)] (a similar finite-size effect at small $L$ can also be seen in Ref.~\onlinecite{shao15}), we can estimate an upper bound of the critical point as $W_c\sim 0.1$ for Gaussian white noise and $W_c\sim 0.05$ for the scatterer ensemble, after which an unambiguous linear scaling of $S\propto L$ starts to show. 
Considering that the CF picture conceptually does not apply to the non-interacting case, the $f=1/2$ metallic critical phase for Gaussian white noise at $W=\infty$, which only exists in the $W=\infty$ limit, is not contradictory with the insulating phase at strong disorder.

\section{Discussion}
\label{discussion}
In this paper, we have tracked the evolution of ground-state entanglement entropy with increasing disorder for electrons with Coulomb interactions at various LLL filling fractions, and estimated the critical points and the length exponents of pertinent disorder-driven phase transitions using a finite-size scaling analysis of the ground state entanglement entropy. Our main results are summarized in Table~\ref{summary}. At $f=1/3,1/5$, and $2/5$, we observe the same feature in the derivative of the entropy with respect to disorder: there is always a pronounced minimum whose position is size independent, but whose magnitude increases markedly with the system size, and is consistent with a divergence in the thermodynamic limit. We consider the location of this minimum as the critical point $W_c$ of the expected transition from a topological FQH state to an insulator. A finite-size scaling analysis of the magnitude of the minimum gives us an estimation of the critical length exponent $\nu$ at these fillings. The values of $\nu$ that we obtain by this method vary by as much as $50\%$ depending on the filling fraction. Moreover, our estimates lie on either side of the conventional bound $\nu\geq 2/d$ for the length exponent for non-topological transitions in disordered systems. This suggests that while our data for the transition from the gapped FQH states (especially for the larger sizes) are quite consistent with finite-size scaling, there are likely corrections to finite-size scaling. These corrections are likely largest for the FQH states with the smallest gaps (like $f = 1/5$). The larger finite-size effects observed at $f=1/5$ and $2/5$ than that at $f=1/3$ may be related to the larger size of composite fermions at these fillings\cite{jainbook}.

\begin{table}
\caption{The estimated zero-disorder gap $\Delta$, the critical disorder strength $W_c$ for Gaussian white noise, and the length exponent $\nu$ extracted from the evolution of entanglement entropy with increasing disorder at various fillings $f$. ``N/A'' means that the value is not available based on our present ED calculation.
The data at $f=1/3$ obtained from the first-order Haldane's pseudopotential interaction are also given in the brackets.}
\begin{ruledtabular}
\begin{tabular}{ccccc}
$f$&$1/5$&$1/3$&$2/5$&$1/2$\\
\hline
$\Delta$&$\sim 0.008$&$\sim 0.06$ [$\sim 0.4$]&$\sim 0.03$&$0$\\
$W_c$&$0.012$&$0.09$ [$0.6$]&$0.06$&$\sim 0.1$\\
$\nu$&$1.3$&$0.9$ [$0.6$]&$0.6$&N/A
\end{tabular}
\end{ruledtabular}
\label{summary}
\end{table}

However, it is noteworthy that all our estimates are very different from $\nu\approx 2.5$ for the localization length at integer plateau transitions\cite{wei88,li05,li09,huo,Huckestein90,chalker88,Slevin09,Lee93,kun96,dnsheng14,chang16}.
It would therefore be of great interest to have experimental estimates of these exponents to see whether they are the same as that for integer plateau transitions. Previous experiments have studied integer quantum Hall plateau transitions by tuning the magnetic field; this may not work for FQH-insulator transition because of the possibility of intervening FQH states. Tuning the disorder, while significantly more challenging, is not out of the question, e.g., by gating a sample, and thereby changing the disorder potential felt by the 2D electron gas. On the numerical side, clearly the best possibilities for improvement remain for FQH states with the largest gaps, which also have the largest $W_c$, as indicated in Table~\ref{summary} by a nearly constant $\Delta/W_c$ that is close to the value found at $f=1/3$ for the Haldane pseudopotential\cite{zhao2016}.

At $f=1/2$, the entropy derivative with respect to disorder has a behavior that is quite size-dependent, varying in a somewhat chaotic manner for the system sizes we are able to study. At this filling, the effective magnetic field for composite fermions vanishes; as a result, there is no gap and the effective length scale (magnetic length) diverges. Consequently, finite-size effects are much more prominent. A similar chaotic behavior is also observed at $f=1/4$ -- another filling fraction where compressible CF liquid is formed at zero disorder. Access to larger systems, probably with the help of more advanced numerical techniques, is needed to verify whether the derivative will become regular when the system is large enough. Even so, extracting $W_c$ and $\nu$ through a finite-size scaling technique could be more complicated because of the different size-dependence of the entanglement entropy for the CF Fermi liquid and the disordered insulator.

Several future directions are suggested by our present work. One is to study the entanglement evolution at $f>1/2$ driven by PH symmetric disorder, and compare the result with that of the PH conjugate at filling $1-f$. It should be noted, however, that because the orbital partition used here cannot distinguish $f$ and $1-f$ in the presence of PH symmetry\cite{sterdyniak2012}, the entanglement computed from a real-space partition\cite{dubail2012,sterdyniak2012} will be necessary in that case.
Another, more interesting topic is to investigate the role of disorder for non-Abelian FQH states, such as those at $f=5/2$ and $12/5$. However, even at zero disorder, these states are more difficult to stabilize than the Abelian states studied in this work. Before considering the disorder effect, we first need to modify the bare Coulomb interaction, for example, by tuning its pseudopotential components or sample thickness\cite{rezayi2000,peterson2008}, to reach robust non-Abelian phases. 

\acknowledgments
We thank Scott Geraedts, Zlatko Papic, and Kun Yang for useful discussions, and an anonymous referee for useful comments. This work was supported by the Department of Energy, Office of Basic Energy Sciences through Grant No.~DE-SC0002140. Z.~L. was also supported by an Alexander von Humboldt Research Fellowship for Postdoctoral Researchers. R.~N.~B. thanks the Aspen Center for Physics for hospitality while this paper was being completed.

\appendix
\section{$f=1/7$}
\label{appa}
Previous studies in clean systems have shown that the proximity to compressible Wigner crystals makes the Coulomb ground states at $f=1/7$ deviate more substantially from the Laughlin model states than the $f=1/3$ and $1/5$ cases\cite{lam84,kun2001,jain2005}. At zero disorder, we indeed find that both the overlap between the Coulomb ground state and the Laughlin model state and the energy gap drop significantly at $N=8$ electrons (Table~\ref{overlap}). The high overlaps for smaller $N$ are then probably because the formation of Wigner crystal, which is sensitive to the sample geometry, is frustrated in these smaller systems on the square torus.
In experiments, the $f=1/7$ FQH effect was also observed only at relatively high temperature\cite{1_7lau,1_7lau2} compared with the $f=1/3$ and $1/5$ cases, which can be explained as the melting of Wigner crystals. Therefore, both numerics and experiments suggest that the $f=1/7$ Coulomb ground state in a large clean system at zero temperature is a compressible Wigner crystal instead of a FQH phase.

\begin{figure}
\centerline{\includegraphics[width=\linewidth]{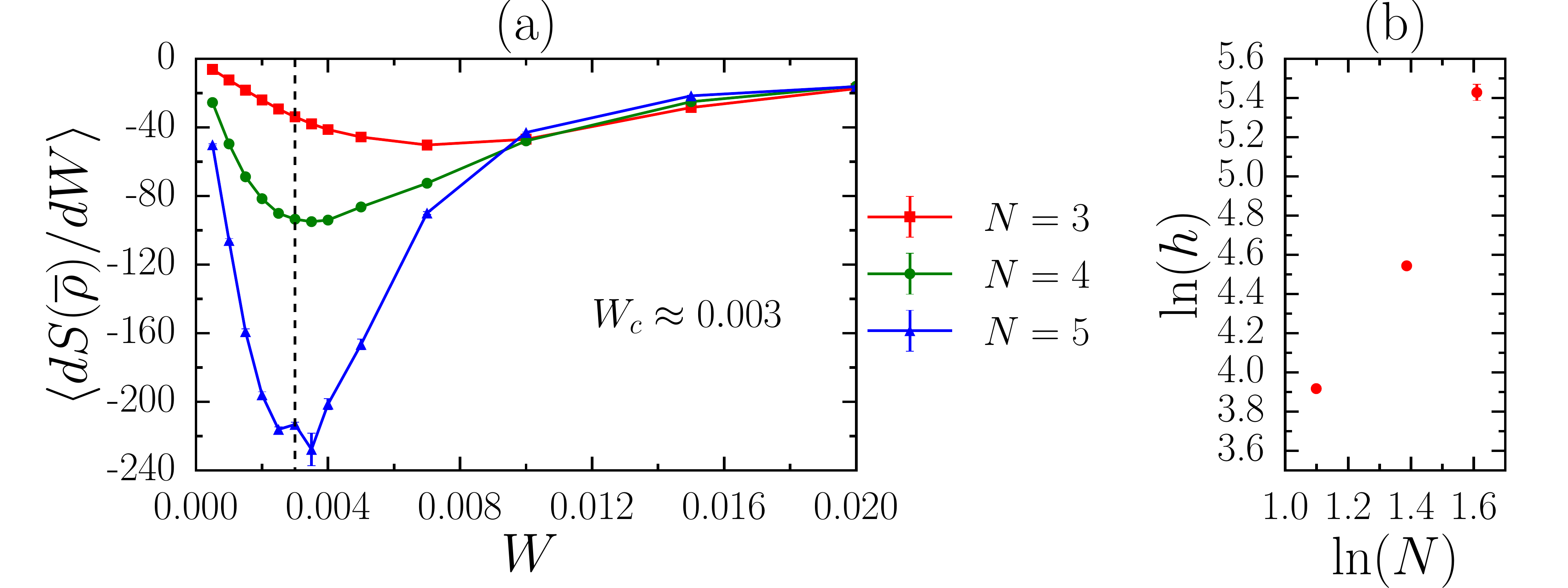}}
\caption{Entanglement entropy measured by $S(\overline{\rho})$ for $N=3-5$ electrons at $f=1/7$. (a) $\langle dS(\overline{\rho})/dW\rangle$ versus
$W$. The vertical dashed line indicates $W=0.003$. (b) The depth of the minimum $h$ versus $N$ on a double logarithmic scale. We averaged $20000$ samples for $N=3$, $10000$ samples for $N=4$, and $2000$ samples for $N=5$ electrons.}
\label{dsdw_1_7}
\end{figure}

In the presence of Gaussian white noise, we track the evolution of the entanglement entropy at $f=1/7$ in a manifold containing the lowest seven eigenstates of the Hamiltonian (\ref{hamil}). Due to the very fast growth of the Hilbert space with increasing electron numbers, we can only efficiently study up to $N=5$ electrons by ED. We find that the behavior of entropy derivative for these very small systems at $f=1/7$ (Fig.~\ref{dsdw_1_7}) are similar to that at $f=1/3$ and $1/5$. There is a pronounced minimum in all $\langle dS(\overline{\rho})/dW\rangle$ curves. Except for the smallest system size $N=3$, the position of this minimum is around a very small value $W_c\approx 0.003$ [Fig.~\ref{dsdw_1_7}(a)], but a reasonable linear fitting of the points in the $\ln h-\ln N$ plot [Fig.~\ref{dsdw_1_7}(b)] is not possible. Nevertheless, this similarity with the $f=1/3$ and $1/5$ cases is probably just a result of the non-vanishing finite-size gaps at zero disorder in these very small systems (Table~\ref{overlap}). Once the compressible Wigner crystal dominates at zero disorder for large enough systems, we expect that the behavior of entanglement entropy would likely become strikingly different.

\begin{figure}
\centerline{\includegraphics[width=\linewidth]{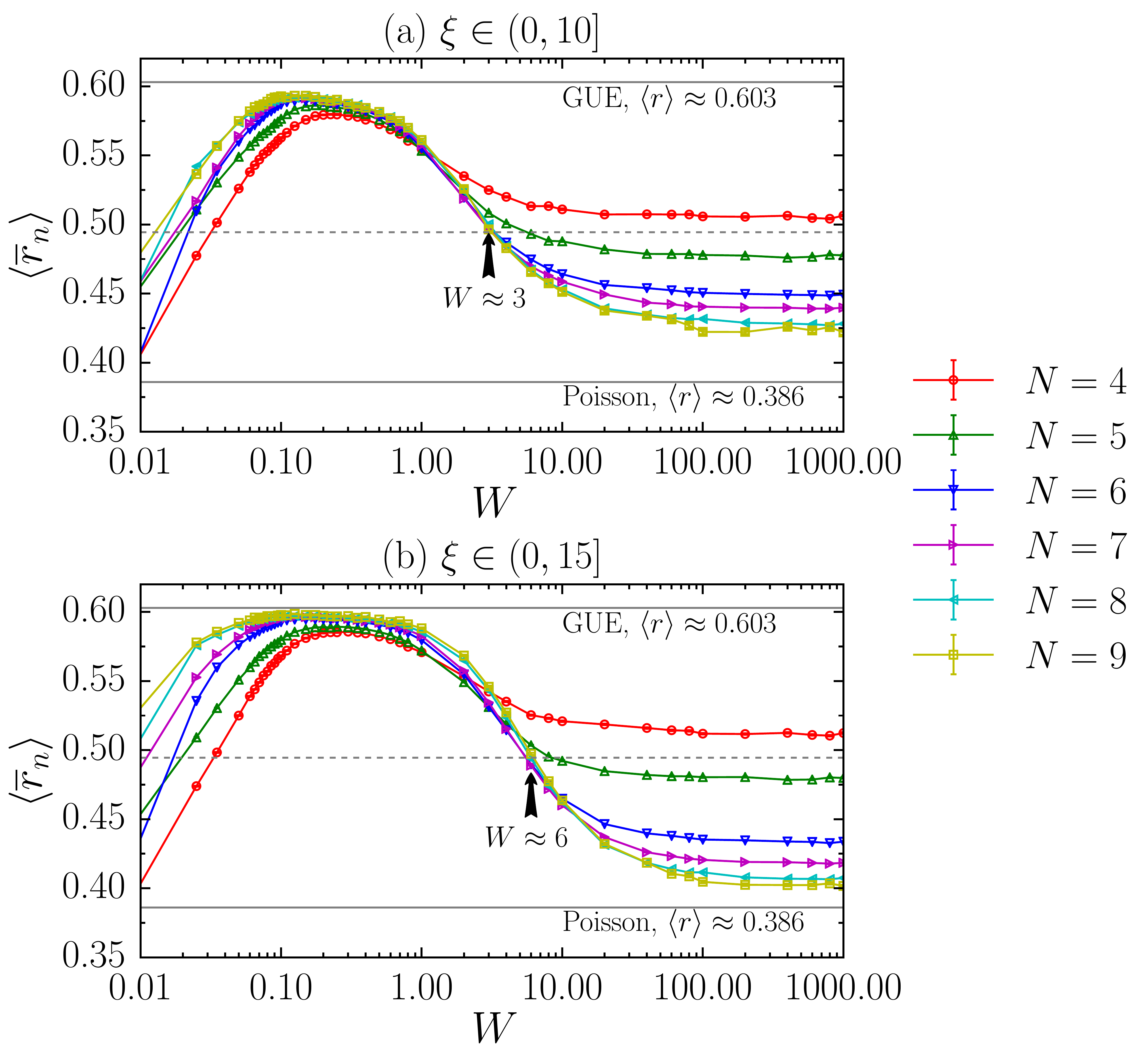}}
\caption{$\langle \overline{r}_n\rangle$ of the ground-state ES for (a) $\xi\in(0,10]$ and (b) $\xi\in(0,15]$. The values of $\langle r\rangle$ for GUE and Poisson distribution are given by solid lines as references. The dashed line corresponds to the middle value $\langle r\rangle\approx 0.495$ between GUE and Poisson. The arrow indicates the transition point of $\langle \overline{r}_n\rangle$ from GUE to Poisson, diagnosed by the crossing of $\langle \overline{r}_n\rangle$ with $\langle r\rangle\approx 0.495$. We averaged $20000$ samples for $N=4-7$, $5000$ samples for $N=8$, and $800$ samples for $N=9$ electrons. }
\label{rxi}
\end{figure}

\section{Entanglement spectrum level statistics}
\label{appb}
Finally, we discuss the level statistics of ground-state ES, i.e., the spectrum of $-\ln\rho_A$, in the presence of Gaussian white noise. In clean systems, each ES level can be labeled by the number of electrons $N_A$ and the total momentum $K_A$ in part $A$\cite{andreas,zhao12}. Disorder breaks the conservation of $K_A$. However, $N_A$ remains a good quantum number, which still allows us to decompose the ES into various $N_A$ sectors. ES levels in different $N_A$ sectors are independent, so putting them together will hide the true level statistics in each $N_A$ sector. Therefore, in the following we will focus on a specific $N_A$ sector, the one with $N_A=\lceil N/2\rfloor$, to study the ES level statistics therein.

In Ref.~\onlinecite{zhao2016}, the ES level statistics were diagnosed using the distribution $P(s)$ of the normalized level spacing $s_n/\langle s_n\rangle$, where $s_n=\xi_n-\xi_{n-1}$ with $\xi_n$'s the unfolded\cite{canali96,nicolas16} ES levels with $N_A=\lceil N/2\rfloor$ sorted in ascending order in each sample. Here we consider the ratio between two consecutive level spacings, i.e., $\langle r_n\rangle$ with $r_n=\min(s_n,s_{n+1})/\max(s_n,s_{n+1})$, as another indicator of the ES level statistics. We first compute $\langle r_n\rangle$ for the ES of each eigenstate of the Hamiltonian (\ref{hamil}) in the ground-state manifold, then further average it over the whole manifold to get the mean $\langle \overline{r}_n\rangle$. In Fig.~\ref{rxi}, we show the evolution of $\langle \overline{r}_n\rangle$ at $f=1/3$ in two windows $\xi\in(0,10]$ and $\xi\in(0,15]$. The results at other fillings are similar.

We indeed observe a transition of $\langle \overline{r}_n\rangle$ from the Gaussian unitary ensemble (GUE) value $\langle r\rangle\approx 0.603$ to the Poisson value $\langle r\rangle\approx 0.386$, which becomes more obvious for larger system sizes. However, contrary to the naive expectation that the ES level statistics might have a dramatic change at the same disorder strength as $W_c\approx 0.09$ where $dS/dW$ diverges, the transition of $\langle \overline{r}_n\rangle$ occurs at a much larger $W_c^\xi$. By assuming that $\langle \overline{r}_n\rangle$ crosses with $\langle r\rangle\approx 0.495$ (i.e., the middle value between GUE and Poisson) at $W=W_c^\xi$, we find $W_c^\xi\approx 3$ for $\xi\in(0,10]$ [Fig.~\ref{rxi}(a)] and $W_c^\xi\approx 6$ for $\xi\in(0,15]$ [Fig.~\ref{rxi}(b)], respectively, which are more than an order of magnitude larger than $W_c\approx 0.09$ indicated by $dS/dW$. In both cases, $W_c^\xi$ almost does not move with increasing system sizes. The smaller $W_c^\xi$ in the $\xi\in(0,10]$ window is consistent with our previous observation for the pseudopotential interaction in Ref.~\onlinecite{zhao2016} that the localization in the ES, reflected by $P(s)$ changing from Gaussian unitary ensemble (GUE) to semi-Poisson and finally to Poisson, is first activated among low levels, then propagates towards higher-$\xi$ region with increasing disorder strength.

Thus, like in Ref.~\onlinecite{zhao2016}, instead of being at (or near) the same disorder strength where $dS/dW$ diverges, the transition point of ES level-spacing statistics from GUE to Poisson is at a very different value of $W$; further, it depends on the choice of the $\xi$ window. Therefore it is not feasible to use this measure of the entanglement spectrum to precisely locate the ground-state phase transition. One may wonder whether using a narrower window above $\xi=0$ can approach $W_c\approx 0.09$. In fact, this is also very difficult, because the ES density of states goes to zero when $\xi\rightarrow 0$\cite{zhao2016}, which means that the number of ES levels in the narrow window is too small to give us a reasonable level statistics.


\begin{thebibliography}{99}

\bibitem{tsui82}
D. C. Tsui, H. L. Stormer, and A. C. Gossard,
\href{https://doi.org/10.1103/PhysRevLett.48.1559}{Phys. Rev. Lett. {\bf 48}, 1559 (1982)}.

\bibitem{laughlin83}
R. B. Laughlin,
\href{https://doi.org/10.1103/PhysRevLett.50.1395}{Phys. Rev. Lett. {\bf 50}, 1395 (1983)}.

\bibitem{hierarchy}
F. D. M. Haldane,
\href{https://doi.org/10.1103/PhysRevLett.51.605}{Phys. Rev. Lett. {\bf 51}, 605 (1983)}.

\bibitem{jain89}
J. K. Jain,
\href{https://doi.org/10.1103/PhysRevLett.63.199}{Phys. Rev. Lett.  {\bf 63}, 199 (1989)}.

\bibitem{moore91}
G. Moore and N. Read,
\href{http://dx.doi.org/10.1016/0550-3213(91)90407-O}{Nucl. Phys. B {\bf 360}, 362 (1991)}.

\bibitem{stormer}
H. L. Stormer,
\href{https://doi.org/10.1103/RevModPhys.71.875}{Rev. Mod. Phys. {\bf 71}, 875 (1999)}.

\bibitem{wen91}
X. G. Wen,
\href{http://dx.doi.org/10.1142/S0217979291001541}{Int. J. Mod. Phys. B, {\bf 05}, 1641 (1991)}.

\bibitem{kitaev03}
A. Y. Kitaev,
\href{http://dx.doi.org/10.1016/S0003-4916(02)00018-0}{Ann. Phys. {\bf 303}, 2 (2003).}

\bibitem{topocompute}
C. Nayak, S. H. Simon, A. Stern, M. Freedman, and S. Das Sarma,
\href{https://doi.org/10.1103/RevModPhys.80.1083}{Rev. Mod. Phys. {\bf 80}, 1083 (2008)}.

\bibitem{laughlin81}
R. B. Laughlin,
\href{https://doi.org/10.1103/PhysRevB.23.5632}{Phys. Rev. B {\bf 23}, 5632(R) (1981)}.

\bibitem{dnsheng03}
D. N. Sheng, X. Wan, E. H. Rezayi, K. Yang, R. N. Bhatt, and
F. D. M. Haldane,
\href{https://doi.org/10.1103/PhysRevLett.90.256802}{Phys. Rev. Lett. {\bf 90}, 256802 (2003)}.

\bibitem{xinwan}
X. Wan, D. N. Sheng, E. H. Rezayi, K. Yang, R. N. Bhatt, and
F. D. M. Haldane,
\href{https://doi.org/10.1103/PhysRevB.72.075325}{Phys. Rev. B {\bf 72}, 075325 (2005)}.

\bibitem{west89}
H. W. Jiang, H. L. Stormer, D. C. Isui, L. N. Pfeiffer, and K. W. West,
\href{https://doi.org/10.1103/PhysRevB.40.12013}{Phys. Rev. B {\bf 40}, 12013(R) (1989)}.

\bibitem{zhang92}
V. Kalmeyer and S.-C. Zhang,
\href{https://doi.org/10.1103/PhysRevB.46.9889}{Phys. Rev. B {\bf 46}, 9889(R) (1992)}.

\bibitem{halperin93}
B. I. Halperin, P. A. Lee, and N. Read,
\href{https://doi.org/10.1103/PhysRevB.47.7312}{Phys. Rev. B {\bf 47}, 7312 (1993)}.

\bibitem{willet93}
R. L. Willett, R. R. Ruel, K. W. West, and L. N. Pfeiffer,
\href{https://doi.org/10.1103/PhysRevLett.71.3846}{Phys. Rev. Lett. {\bf 71}, 3846 (1993)}.

\bibitem{west14}
D. Kamburov, Y. Liu, M. A. Mueed, M. Shayegan, L. N. Pfeiffer, K. W. West, and K. W. Baldwin,
\href{https://doi.org/10.1103/PhysRevLett.113.196801}{Phys. Rev. Lett. {\bf 113}, 196801 (2014)}.

\bibitem{son15}
D. T. Son,
\href{https://doi.org/10.1103/PhysRevX.5.031027}{Phys. Rev. X {\bf 5}, 031027 (2015)}.

\bibitem{shao15}
J. Shao, E.-A. Kim, F. D. M. Haldane, and E. H. Rezayi,
\href{https://doi.org/10.1103/PhysRevLett.114.206402}{Phys. Rev. Lett. {\bf 114}, 206402 (2015)}.

\bibitem{jain15}
A. C. Balram, C. T\ifmmode \mbox{\H{o}}\else \H{o}\fi{}ke, and J. K. Jain,
\href{https://doi.org/10.1103/PhysRevLett.115.186805}{Phys. Rev. Lett. {\bf 115}, 186805 (2015)}.

\bibitem{scottscience}
S. D. Geraedts, M. P. Zaletel, R. S. K. Mong, M. A. Metlitski, A. Vishwanath, and O. I. Motrunich,
\href{http://dx.doi.org/10.1126/science.aad4302}{Science {\bf 352}, 197 (2016)}.

\bibitem{sentil16}
C. Wang and T. Senthil,
\href{https://doi.org/10.1103/PhysRevB.94.245107}{Phys. Rev. B {\bf 94}, 245107 (2016)}.

\bibitem{wang17}
C. Wang, N. R. Cooper, B. I. Halperin, A. Stern,
\href{https://doi.org/10.1103/PhysRevX.7.031029}{Phys. Rev. X {\bf 7}, 031029 (2017)}.

\bibitem{son17}
M. Levin and D. T. Son,
\href{https://doi.org/10.1103/PhysRevB.95.125120}{Phys. Rev. B {\bf 95}, 125120 (2017)}.

\bibitem{matteo17}
M. Ippoliti, S. D. Geraedts, and R. N. Bhatt,
\href{https://doi.org/10.1103/PhysRevB.95.201104}{Phys. Rev. B {\bf 95}, 201104(R) (2017)}.

\bibitem{fqhgammasphere}
M. Haque, O. Zozulya, and K. Schoutens,
\href{https://doi.org/10.1103/PhysRevLett.98.060401}{Phys. Rev. Lett. {\bf 98}, 060401 (2007)}.

\bibitem{hli}
H. Li and F. D. M. Haldane,
\href{https://doi.org/10.1103/PhysRevLett.101.010504}{Phys. Rev. Lett. {\bf 101}, 010504 (2008)}.

\bibitem{nicolas09}
N. Regnault, B. A. Bernevig, and F. D. M. Haldane,
\href{https://doi.org/10.1103/PhysRevLett.103.016801}{Phys. Rev. Lett. {\bf 103}, 016801 (2009)}.

\bibitem{fqhgammatorus}
A. M. L\"auchli, E. J. Bergholtz, and M. Haque,
\href{http://stacks.iop.org/1367-2630/12/i=7/a=075004}{New J. Phys. {\bf 12}, 075004 (2010)}.

\bibitem{andreas}
A. M. L\"auchli, E. J. Bergholtz, J. Suorsa, and M. Haque,
\href{https://doi.org/10.1103/PhysRevLett.104.156404}{Phys. Rev. Lett. {\bf 104}, 156404 (2010)}.

\bibitem{ronny10}
R. Thomale, A. Sterdyniak, N. Regnault, and B. A. Bernevig,
\href{https://doi.org/10.1103/PhysRevLett.104.180502}{Phys. Rev. Lett. {\bf 104}, 180502 (2010)}.

\bibitem{papic11}
Z. Papi\'c, B. A. Bernevig, and N. Regnault,
\href{https://doi.org/10.1103/PhysRevLett.106.056801}{Phys. Rev. Lett. {\bf 106}, 056801 (2011)}.

\bibitem{PES}
A. Sterdyniak, N. Regnault, and B. A. Bernevig,
\href{https://doi.org/10.1103/PhysRevLett.106.100405}{Phys. Rev. Lett. {\bf 106}, 100405 (2011)}.

\bibitem{zhao12}
Z. Liu, E. J. Bergholtz, H. Fan, and A. M. L\"auchli,
\href{https://doi.org/10.1103/PhysRevB.85.045119}{Phys. Rev. B {\bf 85}, 045119 (2012)}.

\bibitem{dubail2012}
J. Dubail, N. Read, and E. H. Rezayi,
\href{https://doi.org/10.1103/PhysRevB.85.115321}{Phys. Rev. B {\bf 85}, 115321 (2012)}.

\bibitem{sterdyniak2012}
A. Sterdyniak, A. Chandran, N. Regnault, B. A. Bernevig, and P. Bonderson,
\href{https://doi.org/10.1103/PhysRevB.85.125308}{Phys. Rev. B {\bf 85}, 125308 (2012)}.

\bibitem{zaletel2013}
M. P. Zaletel, R. S. K. Mong, and F. Pollmann,
\href{https://doi.org/10.1103/PhysRevLett.110.236801}{Phys. Rev. Lett. {\bf 110}, 236801 (2013)}.

\bibitem{wei2015}
W. Zhu, S. S. Gong, F. D. M. Haldane, and D. N. Sheng,
\href{https://doi.org/10.1103/PhysRevLett.115.126805}{Phys. Rev. Lett. {\bf 115}, 126805 (2015)}.

\bibitem{peterson2015}
K. Pakrouski, M. R. Peterson, T. Jolicoeur, V. W. Scarola, C. Nayak, and M. Troyer,
\href{https://doi.org/10.1103/PhysRevX.5.021004}{Phys. Rev. X {\bf 5}, 021004 (2015)}.

\bibitem{estienne2015}
B. Estienne, N. Regnault, and B. A. Bernevig,
\href{https://doi.org/10.1103/PhysRevLett.114.186801}{Phys. Rev. Lett. {\bf 114}, 186801 (2015)}.

\bibitem{friedman11}
B. A. Friedman, G. C. Levine, and D. Luna,
\href{http://stacks.iop.org/1367-2630/13/i=5/a=055006}{New J. Phys. {\bf 13}, 055006 (2011)}.

\bibitem{friedman15}
B. A. Friedman and G. C. Levine,
\href{https://doi.org/10.1142/S0217979215500654}{Int. J. Mod. Phys B {\bf 29}, 1550065 (2015)}.

\bibitem{zhao2016}
Zhao Liu and R. N. Bhatt,
\href{https://doi.org/10.1103/PhysRevLett.117.206801}{Phys. Rev. Lett. {\bf 117}, 206801 (2016)}.

\bibitem{Vnote}
We assume that $V_{\{m_i\}}$ is real, which is true for the isotropic Coulomb interaction. Otherwise, $V_{\{m_i\}}$ will be replaced by $V_{\{m_i\}}^*$ under the PH transform.

\bibitem{haldane85}
F. D. M. Haldane,
\href{https://doi.org/10.1103/PhysRevLett.55.2095}{Phys. Rev. Lett. {\bf 55}, 2095 (1985)}.

\bibitem{zhangyi}
Y. Zhang, T. Grover, A. Turner, M. Oshikawa, and A. Vishwanath,
\href{https://doi.org/10.1103/PhysRevB.85.235151}{Phys. Rev. B {\bf 85}, 235151 (2012)}.

\bibitem{huo}
Y. Huo and R. N. Bhatt,
\href{https://doi.org/10.1103/PhysRevLett.68.1375}{Phys. Rev. Lett. {\bf 68}, 1375 (1992)}.

\bibitem{john11}
J. Schliemann,
\href{https://doi.org/10.1103/PhysRevB.83.115322}{Phys. Rev. B {\bf 83}, 115322 (2011)}.

\bibitem{harris}
A. B. Harris,
\href{http://stacks.iop.org/0022-3719/7/i=17/a=018}{J. Phys. C {\bf 7}, 3082 (1974).}

\bibitem{Chayes86}
J. T. Chayes, L. Chayes, D. S. Fisher, and T. Spencer,
\href{https://doi.org/10.1103/PhysRevLett.57.2999}{Phys. Rev. Lett. {\bf 57}, 2999 (1986)}.

\bibitem{wei1}
W. Zhu, D. N. Sheng, and F. D. M. Haldane,
\href{https://doi.org/10.1103/PhysRevB.88.035122}{Phys. Rev. B {\bf 88}, 035122 (2013)}.

\bibitem{wei2}
W. Zhu, S. S. Gong, F. D. M. Haldane, and D. N. Sheng,
\href{https://doi.org/10.1103/PhysRevLett.112.096803}{Phys. Rev. Lett. {\bf 112}, 096803 (2014)}.

\bibitem{jiang}
H. W. Jiang, R. L. Willett, H. L. Stormer, D. C. Tsui, L. N. Pfeiffer, and K. W. West,
\href{https://doi.org/10.1103/PhysRevLett.65.633}{Phys. Rev. Lett. {\bf 65}, 633 (1990)}.

\bibitem{stormer83}
H. L. Stormer, A. Chang, D. C. Tsui, J. C. M. Hwang, A. C. Gossard, and W. Wiegmann,
\href{https://doi.org/10.1103/PhysRevLett.50.1953}{Phys. Rev. Lett. {\bf 50}, 1953 (1983)}.

\bibitem{Hermanns}
M. Hermanns, J. Suorsa, E. J. Bergholtz, T. H. Hansson, and A. Karlhede,
\href{https://doi.org/10.1103/PhysRevB.77.125321}{Phys. Rev. B {\bf 77}, 125321 (2008)}.

\bibitem{hermann13}
M. Hermanns,
\href{https://doi.org/10.1103/PhysRevB.87.235128}{Phys. Rev. B {\bf 87}, 235128 (2013)}.

\bibitem{anderson}
E. Abrahams, P. W. Anderson, D. C. Licciardello, and T. V. Ramakrishnan,
\href{https://doi.org/10.1103/PhysRevLett.42.673}{Phys. Rev. Lett. {\bf 42}, 673 (1979)}.

\bibitem{Comment}
Numerical results for non-interacting electrons in a random magnetic field [\href{https://doi.org/10.1103/PhysRevB.55.R1922}{Kun Yang and R. N. Bhatt, Phys. Rev. B {\bf55}, R1922 (1997)}] support this view. However, other analytical work [see \href{https://doi.org/10.1103/PhysRevB.49.16609}{A. G. Aronov, A. D. Mirlin and P. W\"olfle, Phys. Rev. B {\bf49}, 16609 (1994)}] suggests an insulating phase with very large localization length at low disorder (even larger than the exponentially large localization lengths for 2D electrons in the absence of a magnetic field\cite{anderson}). Despite of this controversy, which is for quenched (as opposed to fluctuating) random fields, a sharp change in behavior from metallic (or near metallic) to insulating would be expected for the small sizes in our study.

\bibitem{huo93}
Y. Huo, R. E. Hetzel, and R. N. Bhatt,
\href{https://doi.org/10.1103/PhysRevLett.70.481}{Phys. Rev. Lett. {\bf 70}, 481 (1993)}.

\bibitem{wolf06}
M. M. Wolf,
\href{https://doi.org/10.1103/PhysRevLett.96.010404}{Phys. Rev. Lett. {\bf 96}, 010404 (2006)}.

\bibitem{jainbook}
J. K. Jain, {\em Composite Fermions} (Cambridge University Press, Cambridge, 2007).

\bibitem{wei88}
H. P. Wei, D. C. Tsui, M. A. Paalanen, and A. M. M. Pruisken,
\href{https://doi.org/10.1103/PhysRevLett.61.1294}{Phys. Rev. Lett. {\bf 61}, 1294 (1988)}.

\bibitem{li05}
W. Li, G. A. Cs\'athy, D. C. Tsui, L. N. Pfeiffer, and K. W. West,
\href{https://doi.org/10.1103/PhysRevLett.94.206807}{Phys. Rev. Lett. {\bf 94}, 206807 (2005)}.

\bibitem{li09}
W. Li, C. L. Vicente, J. S. Xia, W. Pan, D. C. Tsui, L. N. Pfeiffer, and K. W. West,
\href{https://doi.org/10.1103/PhysRevLett.102.216801}{Phys. Rev. Lett. {\bf 102}, 216801 (2009)}.

\bibitem{Huckestein90}
B. Huckestein and B. Kramer,
\href{https://doi.org/10.1103/PhysRevLett.64.1437}{Phys. Rev. Lett. {\bf 64}, 1437 (1990)}.

\bibitem{chalker88}
J. T. Chalker and P. D. Coddington,
\href{http://stacks.iop.org/0022-3719/21/i=14/a=008}{J. Phys. C {\bf 21}, 2665 (1988)}.

\bibitem{Slevin09}
K. Slevin and T. Ohtsuki,
\href{https://doi.org/10.1103/PhysRevB.80.041304}{Phys. Rev. B {\bf 80}, 041304(R) (2009)}.

\bibitem{Lee93}
D.-H. Lee, Z. Wang, and S. Kivelson,
\href{https://doi.org/10.1103/PhysRevLett.70.4130}{Phys. Rev. Lett. {\bf 70}, 4130 (1993)}.

\bibitem{kun96}
K. Yang and R. N. Bhatt,
\href{https://doi.org/10.1103/PhysRevLett.76.1316}{Phys. Rev. Lett. {\bf 76}, 1316 (1996)}.

\bibitem{dnsheng14}
J. Priest, S. P. Lim, and D. N. Sheng,
\href{https://doi.org/10.1103/PhysRevB.89.165422}{Phys. Rev. B {\bf 89}, 165422 (2014)}.

\bibitem{chang16}
C.-Z. Chang, W. Zhao, J. Li, J. K. Jain, C. Liu, J. S. Moodera, and M. H. W. Chan,
\href{https://doi.org/10.1103/PhysRevLett.117.126802}{Phys. Rev. Lett. {\bf 117}, 126802 (2016)}.

\bibitem{rezayi2000}
E. H. Rezayi and F. D. M. Haldane,
\href{https://doi.org/10.1103/PhysRevLett.84.4685}{Phys. Rev. Lett. {\bf 84}, 4685 (2000)}.

\bibitem{peterson2008}
Michael. R. Peterson, Th. Jolicoeur, and S. Das Sarma,
\href{https://doi.org/10.1103/PhysRevLett.101.016807}{Phys. Rev. Lett. {\bf 101}, 016807 (2008)}.

\bibitem{lam84}
P. K. Lam and S. M. Girvin,
\href{https://doi.org/10.1103/PhysRevB.30.473}{Phys. Rev. B {\bf 30}, 473(R) (1984)}.

\bibitem{kun2001}
K. Yang, F. D. M. Haldane, and E. H. Rezayi,
\href{https://doi.org/10.1103/PhysRevB.64.081301}{Phys. Rev. B {\bf 64}, 081301(R) (2001)}.

\bibitem{jain2005}
C.-C. Chang, G. S. Jeon, and J. K. Jain,
\href{https://doi.org/10.1103/PhysRevLett.94.016809}{Phys. Rev. Lett. {\bf 94}, 016809 (2005)}.

\bibitem{1_7lau}
V. J. Goldman, M. Shayegan, and D. C. Tsui,
\href{https://doi.org/10.1103/PhysRevLett.61.881}{Phys. Rev. Lett. {\bf 61}, 881 (1988)}.

\bibitem{1_7lau2}
W. Pan, H. L. Stormer, D. C. Tsui, L. N. Pfeiffer, K. W. Baldwin, and K. W. West,
\href{https://doi.org/10.1103/PhysRevLett.88.176802}{Phys. Rev. Lett. {\bf 88}, 176802 (2002)}.

\bibitem{canali96}
C. M. Canali,
\href{https://doi.org/10.1103/PhysRevB.53.3713}{Phys. Rev. B {\bf 53}, 3713 (1996)}.

\bibitem{nicolas16}
N. Regnault and R. Nandkishore,
\href{https://doi.org/10.1103/PhysRevB.93.104203}{Phys. Rev. B {\bf 93}, 104203 (2016)}.

\end{thebibliography}
\end{document}